\documentclass[twocolumn,superscriptaddress,eqsecnum,longbibliography,nofootinbib]{revtex4-2}
\usepackage{mathrsfs,amsmath,amsthm,latexsym,amssymb,amsfonts,epsfig,cancel,enumerate,graphicx,subfigure,txfonts}
\usepackage{xcolor}
\definecolor{navy}{RGB}{0,0,150}
\usepackage[colorlinks,linkcolor=navy,citecolor=navy,urlcolor=navy,plainpages=false,pdfstartview=FitH]{hyperref}
\usepackage{appendix}
\allowdisplaybreaks
\newcommand{\dd}{{\rm d}}
\newcommand{\GZU}{School of Physics, Guizhou University, Guiyang 550025, China}
\newcommand{\BNU}{School of Physics and Astronomy, \mbox{Key Laboratory of Multiscale Spin Physics (Ministry of Education)}, Beijing Normal University, Beijing 100875, China}
\newcommand{\UOE}{Department Physik, Institut f\"ur Quantengravitation, Theoretische Physik III, Friedrich-Alexander-Universit\"at Erlangen-N\"urnberg, Staudtstra{\ss}e 7/B2, 91058 Erlangen, Germany}

\begin{document}

\title{Covariant effective spacetimes of spherically symmetric electrovacuum with a cosmological constant}

\author{Jinsong Yang}
\affiliation{\GZU}

\author{Cong Zhang}
\thanks{Corresponding author}
\email{cong.zhang@bnu.edu.cn}
\affiliation{\BNU}
\affiliation{\UOE}

\author{Yongge Ma}
\affiliation{\BNU}

\begin{abstract}
An algebraic framework was introduced in our previous works to address the covariance issue in spherically symmetric effective quantum gravity. This paper extends the framework to the electrovacuum case with a cosmological constant. After analyzing the notion of covariance in the classical theory, we propose an effective Hamiltonian for the electromagnetic field. The effective Hamiltonian together with the effective Hamiltonian constraint of gravity determines an effective dynamical model of gravity coupled to the electromagnetic field. The resulting model is covariant with respect to both the effective metric and the effective vector potential. By solving the equations of motion derived from the effective Hamiltonian constraint, we obtain several quantum-corrected solutions. Notably, some of these solutions reveal quantum gravity effects manifesting not only in spacetime metrics but also in the electromagnetic field. Finally, the covariance of coupling models with general matter fields is discussed.
\end{abstract}

\maketitle

\section{Introduction}

As a well-established theory of gravity, general relativity (GR), predicting the formation of black holes (BHs) and describing the dynamics of the Universe, has successfully passed all experimental and observational tests to date~\cite{Wald:1984rg,Liang:2023ahd}. Nevertheless, the singularities commonly existing according to the singularity theorems~\cite{Penrose:1964wq,Hawking:1970zqf}, due to the geodesic incompleteness of spacetimes, suggest regions where GR breaks down. It is widely expected that a quantum theory of gravity will replace classical GR in these extreme regimes and govern the physics near singularities. As a promising candidate for quantum gravity, loop quantum gravity (LQG) has achieved significant breakthroughs~\cite{Rovelli:2004tv,Thiemann:2007pyv,Thiemann:2002nj,Ashtekar:2004eh,Han:2005km,Giesel:2012ws,Rovelli:2011eq,Perez:2012wv}. In particular, both the canonical and covariant formulations of LQG have been individually developed, providing a full understanding of quantum dynamics~\cite{Thiemann:1996aw,Thiemann:1997rt,Yang:2015zda,Alesci:2015wla,Zhang:2018wbc,Zhang:2019dgi,Engle:2007wy,Freidel:2007py,Alesci:2011ia,Yang:2021den}. Moreover, the concepts and techniques developed in full LQG have been successfully applied to symmetry-reduced models, giving rise to the fields of loop quantum cosmology (LQC) and loop quantum BH (LQBH). Within these frameworks, classical singularities are resolved by some natural scenarios, such as big-bounce cosmology and BH-to-white hole transitions~\cite{Ashtekar:2003hd,Ashtekar:2006rx,Ashtekar:2006wn,Ding:2008tq,Yang:2009fp,Assanioussi:2018hee,Li:2018opr,Ashtekar:2005qt,Gambini:2020nsf}.

Despite these achievements, several issues remain, particularly regarding covariance, which poses a challenge for the effective theory of the quantum symmetry-reduced models. In LQC, covariance is less problematic, as the choice of coordinates adapted to the symmetries of homogeneity and isotropy fully resolves the diffeomorphism constraint. However, in LQBH, while the choice of coordinates adapted to spherical symmetry simplifies the constraint structure, it still leaves the radial component of the diffeomorphism constraint unsolved. Thus, it becomes necessary to consider diffeomorphism transformations along a direction that combines both the temporal and radial coordinates. Consequently, the covariance issue in LQBH models becomes nontrivial and requires careful treatments~\cite{Tibrewala:2013kba,Bojowald:2015zha,Bojowald:2020unm}. This difficulty arises because the effective theory of LQBH is formulated within the Hamiltonian framework, which relies on a $3+1$ decomposition of spacetime.

There are several approaches to preserving covariance in effective LQBH models. One approach is to express the equations in terms of tensors, similar to classical GR. An example of this approach is the construction of the quantum Oppenheimer-Snyder model~\cite{Lewandowski:2022zce,Shi:2024vki}. In this model, a spherically symmetric vacuum is glued to a collapsing ball of spherically symmetric, homogeneous, and pressureless dust. The dynamics of the dust ball is governed by the effective Friedmann equation derived from LQC in~\cite{Ashtekar:2006wn}. It turns out that the exterior spacetime can be uniquely determined using the Israel junction condition~\cite{Israel:1966rt,Poisson:2009pwt}, under the assumption that the exterior region remains stationary. Since the Israel junction condition is a tensor equation, the resulting quantum-corrected exterior spacetime remains covariant. In other words, one can choose different coordinate systems in the exterior region during the matching process and still obtain the same exterior spacetime metric tensor. Another approach is to introduce a specific matter field to fix the gauge of diffeomorphism~\cite{Han:2022rsx,Giesel:2023tsj}. In this approach, different choices of matter fields for gauge fixing can lead to entirely different expressions for the Hamiltonian constraints. Some of these formulations may fail to incorporate key features, such as polymerization, which characterizes LQG. This raises an important question: what kind of matter fields should be selected to ensure that the associated effective Hamiltonian retains these key LQG features? The third approach focuses on constructing an effective Hamiltonian constraint that preserves covariance, as in the classical theory, as done in~\cite{Alonso-Bardaji:2023vtl,Zhang:2024khj,Zhang:2024ney,Belfaqih:2024vfk}. More specifically, the effective Hamiltonian constraint is designed to ensure that the gauge transformations of the metric, generated by the constraints, correspond exactly to its spacetime diffeomorphism transformations. In these works, the approach starts with a modified constraint algebra incorporating quantum gravity effects. The covariance equation for the effective Hamiltonian constraint is then systematically derived based on the necessary and sufficient conditions required to maintain covariance. By solving the covariance equation, one can derive various effective Hamiltonian constraints that correspond to different covariant theories. It is important to note that although these works focus on vacuum gravity, the necessary and sufficient conditions for spacetime covariance are applicable to any matter coupling. Therefore, this approach can be extended to study scenarios beyond the vacuum case, such as the dust-coupling model~\cite{Zhang:2024khj,Zhang:2024ney}. The aim of this paper is to demonstrate that the framework developed in~\cite{Zhang:2024khj,Zhang:2024ney} can also be extended to the electrovacuum setting with a cosmological constant. This extension will ensure covariance not only for the spacetime metric but also for the electromagnetic (EM) field.

When including matter coupling, it is crucial, as emphasized in~\cite{Duque:2024kgu}, to examine the covariance properties of matter fields---particularly when these fields are tensorial. In~\cite{Duque:2024kgu}, the field-strength tensor is chosen as the fundamental variable when imposing the covariance condition, since physical observables must exhibit covariance. While this reasoning is valid, the requirement that physical observables respect covariance can alternatively be implemented by imposing the covariance condition directly on the phase space variables. As we will show, the essential point is that, as gauge fields, these phase space variables are only required to be covariant up to a gauge transformation, which is actually the case in the classical theory. In addition, our approach can be easily generalized to non-Abelian Yang-Mills theory, where the field-strength tensor is no longer gauge invariant.

The remainder of this paper is organized as follows. In Sec.~\ref{sec-II}, we first recall the Hamiltonian formulation of gravity coupled to the EM field in the presentation of a cosmological constant, then analyze the notion of covariance in the classical case, and present its exact solutions of the classical model. In Sec.~\ref{sec-III}, we turn to the effective model of gravity coupled to the EM field. We construct an effective Hamiltonian for the EM field, together with the effective Hamiltonian constraint of gravity, ensuring the covariance of the effective model. Several exact solutions of the effective model are presented. In Sec.~\ref{sec-IV}, the covariance of coupling models with general matter fields is discussed. We show that the resulting spacetime is always covariant for gravity coupled to an arbitrary matter field in effective theory, once the constraint algebra of the coupling model has the similar structure as that of the effective vacuum model. In comparison to gravity, the covariance of matter field needs to be concretely treated. Our results are summarized in Sec.~\ref{sec-V}.

\section{Classical theory}\label{sec-II}

In this section, we will first recall the Hamiltonian formulation of gravity minimally coupled to the EM field in the presentation of a cosmological constant. Building on the Hamiltonian framework, we will check the covariance of the model explicitly. Next, we introduce the reduced formulation of the model with the Gauss constraint solved. Using this reduced formulation, we derive the solutions to the equations of motion (EOM) and examine the covariance of the coupling model based on these solutions.

\subsection{Hamiltonian framework of classical theory}\label{sec-II-A}

Let us focus on the spherically symmetric case. The spacetime is based on a four-dimensional (4D) manifold $\mathcal M\cong\mathbb{R}\times \Sigma$ with $\Sigma\cong \mathcal M_1\times {\mathbb S}^2$. Here $\mathcal M_1$ denotes some one-dimensional manifold which could be, for instance, $\mathbb R_+$, and ${\mathbb S}^2$ is the 2-sphere. Denote the coordinates of the spacetime by $(t,x,\theta,\phi)$. The phase space consists of the canonical variables $(K_1,E^1;K_2,E^2)$ for gravity and $(\Gamma,P_\Gamma)$ for the EM field~\cite{Bojowald:2005cb,Campiglia:2007pr,Zhang:2021xoa,Louko:1996dw,Tibrewala:2012xb}. The nontrivial Poisson brackets between these variables read
\begin{align}
 \left\{K_1(x),E^1(y)\right\}&=2G \delta(x,y),\\
 \left\{K_2(x),E^2(y)\right\}&=G \delta(x,y),\\
 \left\{\Gamma(x),P_\Gamma(y)\right\}&=\delta(x,y),
\end{align}
with $G$ being the Newtonian constant. The physical meaning of the canonical variables becomes evident from the following facts. The variables $E^I$ define the 3D metric
\begin{equation}
 q_{ab}\dd x^a\dd x^b=\frac{(E^2)^2}{E^1}(\dd x+N^x\dd t)^2+E^1\dd\Omega^2,
\end{equation}
where the lowercase latin indices $a,b,\ldots=1,2,3$ stand for the spatial indices and $\dd\Omega^2=\dd\theta^2+\sin^2\theta\dd\phi^2$. The variable $\Gamma$ describes the spatial component of the vector potential of the EM field, namely,
\begin{equation}
 A_a\dd x^a=\Gamma\dd x.
\end{equation}

The dynamics of the coupling model is governed by a set of constraints: the Hamiltonian constraint $H=H^{GR}+H^{EM}$, the diffeomorphism constraint $H_x=H_x^{GR}+H_x^{EM}$, and the Gauss constraint $\mathcal G$ of the EM field. Here the superscripts ``GR'' and ``EM'' stand for gravity (described by GR) and the EM field, respectively. Specifically, we have
\begin{align}
 H^{GR}&:=-\frac{K_1 K_2 \sqrt{E^1}}{G}-\frac{E^2}{2 G \sqrt{E^1}}-\frac{(K_2)^2 E^2}{2G \sqrt{E^1}}+\frac{(\partial_xE^1)^2}{8 G \sqrt{E^1} E^2}\notag\\
 &\hspace{0.5cm}-\frac{\sqrt{E^1} (\partial_xE^1) \partial_xE^2}{2 G (E^2)^2}+\frac{\sqrt{E^1} \partial^2_xE^1}{2GE^2}+\frac{1}{2G}\sqrt{E^1}E^2\Lambda,\label{eq:HGR}\\
 H^{EM}&:=\frac{E^2 P_\Gamma^2}{2(E^1)^{3/2}}+\frac{\sqrt{E^1}}{E^2}\Gamma \mathcal G,\label{eq:HEM}\\
 H_x^{GR}&:=\frac{1}{2G}(2E^2\partial_xK_2-K_1\partial_xE^1),\label{eq:HxGR}\\
 H_x^{EM}&:=-\Gamma \partial_xP_\Gamma,\label{eq:HxEM}\\
 \mathcal G&:=-\partial_xP_\Gamma,\label{eq:gausscons}
\end{align}
where $\Lambda$ denotes the cosmological constant. A straightforward derivation shows that these constraints obey the following algebra:
\begin{subequations}\label{eq:classconstraintalge}
\begin{align}
 \left\{\mathcal G[\beta_1],\mathcal G[\beta_2]\right\}&=0\label{eq:GG1},\\
 \left\{\mathcal G[\beta],H_x[N^x]\right\}&=-\mathcal G[N^x\partial_x\beta],\\
 \left\{\mathcal G[\beta],H[N]\right\}&=-\mathcal G\left[N\frac{\sqrt{E^1}}{E^2}\partial_x\beta\right],\\
 \left\{H_x[N^x_1],H_x[N^x_2]\right\}&=H_x\left[N^x_1\partial_xN^x_2-N^x_2\partial_xN^x_1\right]\label{eq:HxHx},\\
 \left\{H_x[N^x],H[N]\right\}&=H\left[N^x\partial_xN\right]\label{eq:HxH},\\
 \left\{H[N_1],H[N_2]\right\}&=H_x\left[S\left(N^x_1\partial_xN^x_2-N^x_2\partial_xN^x_1\right)\right]\label{eq:HH},
\end{align}
\end{subequations}
with the structure function $S=\frac{E^1}{(E^2)^2}$. Here we introduced the convention $F[g]\equiv\int\dd xF(x)g(x)$ to denote the smeared constraints. Here, the term proportional to the Gauss constraint of the EM field is included as the diffeomorphism constraint of the EM field, as done in~\cite{Tibrewala:2012xb,Alonso-Bardaji:2023niu}. This term ensures that the diffeomorphism constraint generates spatial diffeomorphism transformation for the EM fields. Moreover, the present paper adopts the EM Hamiltonian $H^{EM}$ in Eq.~\eqref{eq:HEM} that differs from the one provided in~\cite{Louko:1996dw,Tibrewala:2012xb} by a term proportional to the Gauss constraint. This additional term is introduced to ensure that the Hamiltonian constraint and the diffeomorphism constraint form a subalgebra, as demonstrated in Eqs.~\eqref{eq:HxHx}--\eqref{eq:HH}, so that the results in our previous work~\cite{Zhang:2024khj} can be applied directly. It is noted that, as a totally constrained system of the first class, these modifications of constraints do not affect the dynamics.

To determine the dynamics in the Hamiltonian framework, we need to choose a lapse function $N$, a shift vector $N^x\partial_x$, and a Lagrangian multiplier $\Psi$ to define a Hamiltonian
\begin{equation}
 \mathbb{H}=H[N]+H_x[N^x]+\mathcal G[\Psi].
\end{equation}
Then, Hamilton's equations, as the EOM, read
\begin{align}
 \dot{K_1}&:=\left\{K_1,{\mathbb H}\right\}=\left\{K_1,{\mathbb H}^{GR}\right\}-N\frac{3GE^2}{2\left(E^1\right)^{5/2}}P_\Gamma^2+N\frac{G}{\sqrt{E^1}E^2}\Gamma\mathcal G\label{eq:dotK1},\\
 \dot{K_2}&:=\left\{K_2,{\mathbb H}\right\}=\left\{K_2,{\mathbb H}^{GR}\right\}+N\frac{G}{2\left(E^1\right)^{3/2}}P_\Gamma^2-N\frac{G\sqrt{E^1}}{(E^2)^2}\Gamma\mathcal G,\\
 \dot{E^1}&:=\left\{E^1,{\mathbb H}\right\}=\left\{E^1,{\mathbb H}^{GR}\right\},\\
 \dot{E^2}&:=\left\{E^2,{\mathbb H}\right\}=\left\{E^2,{\mathbb H}^{GR}\right\},\\
 \dot{\Gamma}&:=\left\{\Gamma,{\mathbb H}\right\}=\partial_x\Psi+\partial_x\left[\left(N\frac{\sqrt{E^1}}{E^2}+N^x\right)\Gamma\right]+N\frac{E^2}{\left(E^1\right)^{3/2}}P_\Gamma,\label{eq:gammadot}\\
 \dot{P_\Gamma}&:=\left\{P_\Gamma,{\mathbb H}\right\}=-\left(N\frac{\sqrt{E^1}}{E^2}+N^x\right)\mathcal G,\label{eq:pgammadot}
\end{align}
where
\begin{align}
 {\mathbb H}^{GR}:=H^{GR}[N]+H^{GR}_x[N^x],
\end{align}
stands for the gravitational Hamiltonian. Solving Hamilton's equations together with the constraint equations $H=0$, $H_x=0$ and $\mathcal G=0$, we get the solutions $E^I(t,x)$, $K_I(t,x)$, $\Gamma(t,x)$, and $P_\Gamma(t,x)$ as fields on $\mathcal M$. Denote the spacetime indices by the Greek letters $\rho,\sigma,\ldots=0,1,2,3$. Then, the 4D metric $g_{\rho\sigma}$ and the 4D vector potential $A_{\rho}$ are given by~\cite{Louko:1996dw,Tibrewala:2012xb}
\begin{align}
 \dd s^2&=-N^2\dd t^2+\frac{(E^2)^2}{E^1}(\dd x+N^x\dd t)^2+E^1\dd \Omega^2,\label{eq:lineelem}\\
 A_{\rho}\dd x^{\rho} &=\Phi \dd t+\Gamma \dd x,\label{eq:vectorpotential}
\end{align}
with $\Phi$ relating to $\Psi$ and $\Gamma$ by
\begin{equation}\label{eq:tildePhiPhi}
 \Phi=\Psi+\left(N\frac{\sqrt{E^1}}{E^2}+N^x\right)\Gamma.
\end{equation}

\subsection{Covariance of classical theory}\label{sec-II-B}

In this subsection, we will demonstrate the covariance of the classical theory within the Hamiltonian framework algebraically.

It is observed that $H_x[N^x]$ generates the diffeomorphism transformation along the vector field $N^x\partial_x$. Therefore, Hamilton's equations \eqref{eq:dotK1}--\eqref{eq:pgammadot} can be rewritten by the Lie derivatives as
\begin{equation}\label{eq:EOMlie}
 \begin{aligned}
 \mathcal L_{\mathfrak N}K_I=&\left\{K_I,H[N]\right\},\\
 \mathcal L_{\mathfrak N}E^I=&\left\{E^I,H[N]\right\},\\
 \mathcal L_{\mathfrak N}\Gamma=&\left\{\Gamma,H[N]+\mathcal G[\Psi]\right\},\\
 \mathcal L_{\mathfrak N}P_\Gamma=&\left\{P_\Gamma,H[N]+\mathcal G[\Psi]\right\},
 \end{aligned}
\end{equation}
where $I=1,2$, $\mathfrak N=\partial_t-N^x\partial_x$, and $\mathcal L_{\mathfrak N}$ denotes the Lie derivative along the vector $\mathfrak N$ on $\mathcal M$. The fields $K_1$, $E^2$, $\Gamma$ are scalar densities with weight $1$, and $K_2$, $E^1$, and $P_\Gamma$ are scalars, which is relevant for calculating the Lie derivative. Note that by converting Eqs.~\eqref{eq:dotK1}--\eqref{eq:pgammadot} into Eq.~\eqref{eq:EOMlie}, we indeed establish the relationship between the evolution of a point in phase space and the corresponding evolution of the fields in spacetime.

Now, let us consider a gauge transformation generated by $H[\alpha N]+H_x[\beta^x]+\mathcal G[\alpha\Psi+\lambda]$, for arbitrary smeared functions $\alpha$ and $\lambda$, and an arbitrary smeared vector field $\beta^x\partial_x$. To show the covariance of the coupling model, we need to demonstrate that the gauge transformations in phase space correspond to the diffeomorphism transformations in spacetime. More precisely, we will prove
\begin{align}
 \delta g_{\rho\sigma}&=\mathcal L_{\alpha \mathfrak N+\beta}g_{\rho\sigma},\label{eq:covariant-g-class}\\
 \delta A_{\rho}&=\mathcal L_{\alpha \mathfrak N+\beta}A_{\rho}+\partial_\rho\lambda.\label{eq:covariant-A-class}
\end{align}
Note that in Eq.~\eqref{eq:covariant-A-class}, the extra term $\partial_\rho \lambda$, representing an infinitesimal U(1) gauge transformation, arises because $A$ is a gauge field with U(1) gauge symmetry.

Under the gauge transformation, the phase space variables $W\equiv K_I,E^I,\Gamma,P_\Gamma$ transform as
\begin{align}\label{eq:gauge}
 W\rightarrow \tilde{W}\equiv W+\epsilon \delta W+O(\epsilon^2),
\end{align}
where $\epsilon$ denotes a small constant that characterizes the parameter of one parameter diffeomorphism generalized by $H[\alpha N]+H_x[\beta^x]+\mathcal G[\alpha\Psi+\lambda]$, and
\begin{align}\label{eq:deltaW}
 \delta W:=\left\{W,H[\alpha N]+H_x[\beta^x]+\mathcal G[\alpha\Psi+\lambda]\right\}.
\end{align}
As explicitly computed in Appendix~\ref{app:A}, this yields the following transformation laws:
\begin{equation}\label{eq:deltaEI}
 \begin{aligned}
 \delta E^1&=\mathcal L_{\alpha\mathfrak N+\beta}E^1,\\
 \delta E^2&=\mathcal L_{\alpha\mathfrak N+\beta}E^2-E^2\mathfrak N^\rho\partial_\rho\alpha,\\
 \delta\Gamma=&\mathcal L_{\alpha \mathfrak N+\beta}\Gamma+\left(N\frac{\sqrt{E^1}}{E^2}\Gamma+\Psi\right)\partial_x\alpha-\Gamma\mathfrak N^{\rho}\partial_{\rho}\alpha+\partial_x\lambda.
 \end{aligned}
\end{equation}

Since the expression of the metric $g_{\rho\sigma}$ and the EM field $A_\rho$ involves also the Lagrangian multiplier $N$, $N^x$, and $\Psi$, which could be phase space independent, a subtle issue arises regarding how these phase-space-independent variables transform under gauge transformations. This issue has been thoroughly discussed in~\cite{Zhang:2024khj,Zhang:2024ney}, where we show that the transformation laws of the phase-space-independent variables are determined by the requirement that $\tilde{W}$, defined in Eq.~\eqref{eq:gauge}, satisfies the equation of motion \eqref{eq:EOMlie} associated with the transformed Lagrange multipliers $N + \epsilon \delta N$, $N^x + \epsilon \delta N^x$, and $\Psi + \epsilon \delta \Psi$. As explicitly derived in Appendix \ref{app:A}, this requirement leads to the following transformation rules:
\begin{equation}\label{eq:dNNxEM}
 \begin{aligned}
 \delta N=&\mathcal L_{\alpha\mathfrak N+\beta}N+N\mathfrak N^\rho\partial_\rho\alpha,\\
 \delta N^x=&-N^2S \partial_x\alpha-(\mathcal L_{\beta}\mathfrak N)^x,\\
 \delta\Psi=&-N\sqrt{S}(\Psi\partial_x\alpha+\partial_x\lambda)+\beta^x\partial_x\Psi+\mathcal L_{\mathfrak N}(\alpha\Psi+\lambda).
 \end{aligned}
\end{equation}
Combining Eqs.~\eqref{eq:deltaEI} and \eqref{eq:dNNxEM}, we can directly show \eqref{eq:covariant-g-class} and \eqref{eq:covariant-A-class}.

\subsection{Reduced formulation}\label{sec-II-C}

In this subsection, we reduce the degrees of freedom by solving the Gauss constraint, which simplifies the process of obtaining solutions in the next subsection.

Combining the evolution equation \eqref{eq:pgammadot} with the constraint equation $\mathcal G=G^{EM}=0$ leads to that $P_\Gamma$ is a constant~\cite{Louko:1996dw}
\begin{align}
 P_\Gamma=Q,
\end{align}
which is independent of both $t$ and $x$. Moreover, $H_x^{EM}$ vanishes due to $H^{EM}_x$ being proportional to $\mathcal G$. Notice that both the remaining constraints and the right-hand sides of Hamilton's equations are independent of $\Gamma$. Hence, we can choose the partial gauge fixing~\cite{Gambini:2014qta}
\begin{align}\label{eq:partialGF}
 \Gamma=0.
\end{align}
As a result, Eq.~\eqref{eq:tildePhiPhi} reduces to
\begin{align}\label{eq:Psi_Phi_class}
 \Psi=\Phi.
\end{align}
Equation~\eqref{eq:gammadot} reads
\begin{align}
 0&=\partial_x\Psi+N\frac{E^2}{\left(E^1\right)^{3/2}}Q.\label{eq:tildePhi}
\end{align}
Now, the pair of $(\Gamma,P_\Gamma)$ is not dynamical and thus can be dropped out from the canonical variables. Therefore, after solving the Gauss constraint equation $\mathcal G=0$ and imposing the partial gauge fixing \eqref{eq:partialGF}, the phase space of the coupling model consists of $(K_1,E^1;K_2,E^2)$, and the Hamiltonian constraint $H$ and the diffeomorphism constraint $H_x$ reduce to
\begin{align}
 \bar{H}&=H^{GR}+H^Q,\label{eq:barHclass}\\
 \bar{H}_x&=H_x^{GR}\label{eq:barHxclass},
\end{align}
where
\begin{align}\label{eq:HQ-clas}
 H^Q\equiv\left.H^{EM}\right|_{P_\Gamma=Q}=\frac{E^2 Q^2}{2(E^1)^{3/2}}.
\end{align}
The resulting constraint algebra reads
\begin{subequations}\label{eq:constraint-algebra-GR-reduced}
\begin{align}
 \left\{\bar{H}_x(N^x_1),\bar{H}_x(N^x_2)\right\}&=\left\{H_x^{GR}(N^x_1),H_x^{GR}(N^x_2)\right\}\notag\\
 &=\bar{H}_x(N^x_1\partial_xN^x_2-N^x_2\partial_xN^x_1),\label{eq:algebra1}\\
 \left\{\bar{H}_x(N^x],\bar{H}(N)\right\}&=\bar{H}(N^x\partial_xN),\\
 \left\{\bar{H}(N_1),\bar{H}(N_2)\right\}&=\left\{H^{GR}(N_1),H^{GR}(N_2)\right\}\notag\\
 &=\bar{H}_x\left[S\left(N_1\partial_xN_2-N_2\partial_xN_1\right)\right].\label{eq:barHcommu}
\end{align}
\end{subequations}
In the first step in Eq.~\eqref{eq:barHcommu}, we notice that $H^{GR}$ does not involve the derivatives of $K_I$, while $H^Q$ is only a function of $E^I$ but does not involve the derivatives of $E^I$. Thus, $H^Q$ does not contribute to the Poisson bracket between the smeared Hamiltonian constraints. It is worthwhile to note that the constraint algebra \eqref{eq:constraint-algebra-GR-reduced} for the coupling model is precisely the same as the one in the vacuum case. Then, the reduced Hamiltonian
\begin{align}
 \bar{\mathbb H}&:=\int\dd x\left(N\bar{H}+N^x\bar{H}_x\right),
\end{align}
determines corresponding Hamilton's equations
\begin{align}
 \dot{K_1}&:=\left\{K_1,\bar{\mathbb H}\right\}=\left\{K_1,{\mathbb H}^{GR}\right\}-N\frac{3GE^2}{2\left(E^1\right)^{5/2}}Q^2,\label{eq:barK1dot}\\
 \dot{K_2}&:=\left\{K_2,\bar{\mathbb H}\right\}=\left\{K_2,{\mathbb H}^{GR}\right\}+N\frac{G}{2\left(E^1\right)^{3/2}}Q^2,\\
 \dot{E^1}&:=\left\{E^1,\bar{\mathbb H}\right\}=\left\{E^1,{\mathbb H}^{GR}\right\},\\
 \dot{E^2}&:=\left\{E^2,\bar{\mathbb H}\right\}=\left\{E^2,{\mathbb H}^{GR}\right\}.\label{eq:barE2dot}
\end{align}

\subsection{Classical solutions}\label{sec-II-D}

In this subsection, by solving Eq.~\eqref{eq:tildePhi}, the constraint equations $\bar{H}=0$ and $\bar{H}_x=0$ and Hamilton's equations \eqref{eq:barK1dot}--\eqref{eq:barE2dot} under different gauge fixings, we will obtain the corresponding solutions. Based on the solutions, it is more obvious to understand the covariance from the viewpoint of coordinate transformation.

To this end, let us choose the areal gauge~\cite{Bojowald:2006ip,Tibrewala:2012xb,Gambini:2014qta}
\begin{align}\label{eq:arealgauge}
 E^1=x^2.
\end{align}
Then, the diffeomorphism constraint equation $\bar{H}_x=0$ yields
\begin{align}\label{eq:solvediff}
 K_1=\frac{E^2\partial_xK_2}{x}.
\end{align}
The areal gauge \eqref{eq:arealgauge} allows us to find the stable solution that satisfies the following stationary condition:
\begin{align}
 0&=\dot{E^1}=\left\{E^1,\bar{H}[N]+\bar{H}_x[N^x]\right\},\\
 0&=\dot{E^2}=\left\{E^2,\bar{H}[N]+\bar{H}_x[N^x]\right\}.
\end{align}
This condition can be used to determine the explicit forms $N$ and $N^x$ as
\begin{align}
 N&=\frac{x}{E^2},\label{eq:NExp}\\
 N^x&=-\frac{K_2}{E^2}x.\label{eq:NxExp}
\end{align}
Inserting Eqs.~\eqref{eq:arealgauge} and \eqref{eq:solvediff} into Eq.~\eqref{eq:barHclass} yields
\begin{align}
 \bar{H}=-\frac{E^2}{x}\partial_x\left\{\frac{x}{2G}\left[1+\frac{GQ^2}{x^2}-\frac{\Lambda}{3}x^2+x^2\left(\frac{(K_2)^2}{x^2}-\frac{1}{(E^2)^2}\right)\right]\right\}.
\end{align}
Hence, the Hamiltonian constraint equation $\bar{H}=0$ implies that the term in the derivative is a constant $M$, i.e.,
\begin{align}\label{eq:Mconst}
 \frac{x}{2G}\left[1+\frac{GQ^2}{x^2}-\frac{\Lambda}{3}x^2+x^2\left(\frac{(K_2)^2}{x^2}-\frac{1}{(E^2)^2}\right)\right]=M.
\end{align}
Furthermore, putting Eqs.~\eqref{eq:arealgauge} and \eqref{eq:NExp} into Eq.~\eqref{eq:tildePhi} yields
\begin{align}
 0&=\partial_x\Psi+\frac{Q}{x^2}.
\end{align}
It implies
\begin{align}
 \Phi=\Psi=\frac{Q}{x},
\end{align}
where we used Eq.~\eqref{eq:Psi_Phi_class}. Thus, the vector potential \eqref{eq:vectorpotential} reduces to
\begin{align}\label{eq:vectorpotentialSol}
 A_\rho\dd x^\rho&=\frac{Q}{x} \dd t.
\end{align}

In what follows, we aim to obtain the explicit solutions in different coordinates. On the one hand, we take the Schwarzschild gauge
\begin{align}\label{eq:SchGauge}
 N^x=0.
\end{align}
Inserting Eq.~\eqref{eq:SchGauge} into Eq.~\eqref{eq:NxExp} yields
\begin{align}\label{eq:K2value}
 K_2&=0.
\end{align}
Putting Eq.~\eqref{eq:K2value} into Eq.~\eqref{eq:Mconst}, we have
\begin{align}\label{eq:E2value}
 E^2&=\pm\frac{x}{\sqrt{f_{\rm cl}(x)}},
\end{align}
where
\begin{align}
 f_{\rm cl}(x)=1-\frac{2GM}{x}+\frac{GQ^2}{x^2}-\frac{\Lambda}{3}x^2.
\end{align}
Putting Eq.~\eqref{eq:E2value} into Eq.~\eqref{eq:NExp} yields
\begin{align}
 N^2=f_{\rm cl}(x).
\end{align}
Thus, we obtain the classical solution in Schwarzschild coordinates with the line element \eqref{eq:lineelem} being
\begin{align}\label{eq:lineelem-Sch}
 \dd s^2=-f_{\rm cl}(x)\dd t^2+f_{\rm cl}(x)^{-1}\dd x^2+x^2\dd \Omega^2.
\end{align}
It is the Reissner-Nordstr\"om (RN) solution with a cosmological constant as expected. On the other hand, we take the Painlev\'e-Gullstrand (PG) gauge
\begin{align}\label{eq:PG-gauge}
 E^2=x.
\end{align}
According to Eq.~\eqref{eq:NExp}, the PG gauge \eqref{eq:PG-gauge} results in
\begin{align}\
 N=1.
\end{align}
Inserting Eq.~\eqref{eq:PG-gauge} into Eqs.~\eqref{eq:NxExp} and \eqref{eq:Mconst}, we get
\begin{align}
 N^x&=\pm\sqrt{1-f_{\rm cl}(x)}.
\end{align}
Then, the solution in PG coordinates is obtained, corresponding to the line element \eqref{eq:lineelem} taking
\begin{align}\label{eq:lineelem-PG}
 \dd s^2&=-\dd \tau^2+\left[\dd x\pm\sqrt{1-f_{\rm cl}(x)}\,\dd \tau\right]^2+x^2\dd \Omega^2,
\end{align}
where $\tau$ is used to denote the time coordinate in the PG gauge.

These two solutions \eqref{eq:lineelem-Sch} and \eqref{eq:lineelem-PG} are related by the coordinate transformation
\begin{align}
 \dd t=\dd \tau\mp\frac{\sqrt{1-f_{\rm cl}(x)}}{f_{\rm cl}(x)}\,\dd x,
\end{align}
reflecting the fact that the spacetime is covariant. Similarly, the vector potential \eqref{eq:vectorpotentialSol} in these two coordinates can be related by
\begin{align}
 A_\rho\dd x^\rho&=\frac{Q}{x}\dd t=\frac{Q}{x}\dd \tau\mp\dd Y(x),
\end{align}
where $Y$ is the solution to the ordinary differential equation
\begin{equation}
 \frac{\dd Y}{\dd x}=\frac{Q}{x}\frac{\sqrt{1-f_{\rm cl}(x)}}{f_{\rm cl}(x)}.
\end{equation}
Hence, two vector potentials obtained in different gauges differ only in the gauge term $\dd Y(x)$, implying the covariance of the EM field.

\section{Effective theory}\label{sec-III}

In the above section, we introduce how the EM field is coupled to gravity within the Hamiltonian framework of the classical GR. Additionally, the general covariance of the classical model was examined with detailed calculations. Now let us turn to the effective model of the EM field coupled to the effective quantum gravity proposed by~\cite{Zhang:2024khj,Zhang:2024ney}, and show its covariance.

\subsection{Covariance of effective theory}\label{sec-III-A}

In~\cite{Zhang:2024khj,Zhang:2024ney}, although the focus is on the vacuum case, the conclusions can be extended to scenarios with matter coupling. To provide a comprehensive perspective, we first introduce these conclusions in a general setting without excluding matter coupling.

In these works, the diffeomorphism constraint $H_x$ that generates the spatial diffeomorphism transformation takes its classical form, while the Hamiltonian constraint is modified into an effective one, denoted by $H_{\rm eff}$, with quantum gravity effects included. They are assumed to satisfy the algebraic relation
\begin{subequations}\label{eq:constraint-algebra-GR-eff}
\begin{align}
 \left\{H_x[N^x_1],H_x[N^x_2]\right\}&=H_x\left[N^x_1\partial_xN^x_2-N^x_2\partial_xN^x_1\right]\label{eq:HxHxeff},\\
 \left\{H_x[N^x],H_{\rm eff}[N]\right\}&=H_{\rm eff}\left[N^x\partial_xN\right],\\
 \left\{H_{\rm eff}[N_1],H_{\rm eff}[N_2]\right\}&=H_x\left[\mu S\left(N^x_1\partial_xN^x_2-N^x_2\partial_xN^x_1\right)\right],\label{eq:HHeff}
\end{align}
\end{subequations}
where $\mu$, which depends only on the gravitational degrees of freedom, appears in Eq.~\eqref{eq:HHeff} as a result of quantum gravity effects. To match this modified constraint algebra, the classical metric is replaced with an effective metric $g_{\rho\sigma}^{(\mu)}$, given by the line element
\begin{equation}\label{eq:lineelem-eff}
 \dd s^2=-N^2\dd t^2+\frac{(E^2)^2}{\mu E^1}(\dd x+N^x\dd t)^2+E^1\dd \Omega^2.
\end{equation}
It should be mentioned that $\mu$ may change sign, as observed in some black hole models~\cite{Zhang:2024khj}. At first glance, such a sign change may seem problematic, as it would suggest that a spacetime region with signature $(-,-,+,+)$ must be smoothly joined to a standard Lorentzian region. However, as discussed in our previous work~\cite{Zhang:2024khj}, this situation does not actually arise. The spacetime points where $\mu = 0$ correspond to coordinate singularities. To extend the spacetime across $\mu = 0$, one must adopt a different coordinate system that remains regular at these points. Using such coordinates, one will glue the region with $\mu > 0$ to the original Lorentzian region in a manner that yields a spacetime symmetric with respect to the hypersurface $\mu = 0$, without introducing any region where $\mu < 0$ (see also Secs.~\ref{sec:metricIIspa} and \ref{sec:propertyIII} for concrete examples related to this issue).\footnote{A similar scenario was previously encountered in certain quantum cosmological models (see, e.g.,~\cite{Cailleteau:2011kr}), where the metric, at least in specific cases, takes the form $\dd s^2 = -f \dd\eta^2 + g \delta_{ab} \dd x^a \dd x^b.$ Since $f$ is not positive definite, this may suggest a signature change to $(+,+,+,+)$ at $f = 0$. As discussed in the main text, such predictions require more careful and nuanced analysis due to the degeneracy of the metric at $f = 0$, even though it remains unclear whether these points correspond to coordinate singularities.}

With these definitions and assumptions, it turns out in~\cite{Zhang:2024khj,Zhang:2024ney} that the metric $g_{\rho\sigma}^{(\mu)}$ is covariant, meaning that it transforms as
\begin{align}\label{eq:covariant-g-eff}
 \delta g_{\rho\sigma}^{(\mu)}&=\mathcal L_{\alpha \mathfrak N+\beta}g_{\rho\sigma}^{(\mu)},
\end{align}
where $\delta g_{\rho\sigma}^{(\mu)}$ is the infinitesimal gauge transformation generated by $H_{\rm eff}[\alpha N]+H_x[\beta^x]$, if and only if the following two conditions are satisfied:
\begin{enumerate}[(i)]
\item $H_{\rm eff}$ is independent of the derivatives of $K_1$;
\item For all phase-space-independent functions $\alpha(x)$ and $N(x)$, the following condition holds:
\begin{equation}
 \{\mu(x)S(x),H_{\rm eff}[\alpha N]\}=\alpha(x)\{\mu(x)S(x),H_{\rm eff}[N]\}.
\end{equation}
\end{enumerate}

In the vacuum case, $H_{\rm eff}$ is just the effective Hamiltonian constraint of gravity, denoted by $H_{\rm eff}^{GR}$. It is noted by~\cite{Zhang:2024khj,Zhang:2024ney} that $H_{\rm eff}^{GR}$, as a scalar density, must take the form $E^2F$ where $F$ is a function of some basic scalars constructed from by $E^I$, $K_I$, and their derivatives. With restricting $F$ to depend on the same basic scalars $s_a$ ($a=1,\ldots,5$) as in the classical theory, namely,
\begin{equation}
 \begin{aligned}
 &s_1=E^1,\qquad\ s_2=K_2,\qquad s_3=\frac{K_1}{E^2},\\
 &s_4=\frac{\partial_xE^1}{E^2},\qquad s_5=\frac{1}{E^2}\partial_x\left(\frac{\partial_xE^1}{E^2}\right),
 \end{aligned}
\end{equation}
it is derived from the above two conditions (i) and (ii) that $H_{\rm eff}^{GR}$ must be given by the following equations~\cite{Zhang:2024khj,Zhang:2024ney}:
\begin{equation}\label{eq:covariance-Heff}
 \begin{aligned}
 H_{\rm eff}^{GR}=&-2E^2\left[\partial_{s_1}M_{\rm eff}+\frac{\partial_{s_2}M_{\rm eff}}{2}s_3+\frac{\partial_{s_4}M_{\rm eff}}{s_4}s_5+\mathcal R(s_1,M_{\rm eff})\right],
 \end{aligned}
\end{equation}
where $\mathcal R$ is an arbitrary function of $s_1$ and $M_{\rm eff}(s_1, s_2, s_4)$, and $M_{\rm eff}(s_1, s_2, s_4)$ satisfies the following equations evolving a function $\mu(s_1,s_2,s_4)$:
\begin{subequations}\label{eq:covariance-mu}
\begin{align}
 &\frac{\mu s_1 s_4}{4G^2}=\left(\partial_{s_2}M_{\rm eff}\right)\partial_{s_2}\partial_{s_4}M_{\rm eff}-\left(\partial_{s_4}M_{\rm eff}\right)\partial_{s_2}^2M_{\rm eff},\label{eq:covarianceequationgeneral1}\\
 &\left(\partial_{s_2}\mu\right)\partial_{s_4}M_{\rm eff}-\left(\partial_{s_2}M_{\rm eff}\right)\partial_{s_4}\mu=0.\label{eq:covarianceequationgeneral2}
\end{align}
\end{subequations}
Equation~\eqref{eq:covarianceequationgeneral2} implies
\begin{align}\label{eq:mus1Meff}
 \mu=\mu(s_1,M_{\rm eff}).
\end{align}
See~\cite{Zhang:2025ccx} for more details on general solutions to Eq.~\eqref{eq:covariance-mu}. It is easy to show that the classical Hamiltonian constraint \eqref{eq:HGR} can be recovered if we set the free function $\mathcal R$ to be
\begin{align}
 \mathcal R_{\rm cl}&=-\frac{\Lambda}{4G}\sqrt{s_1},
\end{align}
and take
\begin{align}\label{eq:M-cl}
 M_{\rm eff}\equiv M_{\rm cl}&=\frac{\sqrt{s_1}}{2G}\left[1+(s_2)^2-\frac{(s_4)^2}{4}\right],
\end{align}
which is a solution to Eq.~\eqref{eq:covariance-mu} with $\mu=1$.

In this paper, \text{from this point onward}, to incorporate the quantum gravity effects, we will take
\begin{align}\label{eq:Reff}
 \mathcal R&=-\mathcal X(\mu)\frac{\Lambda}{4G}\sqrt{s_1},
\end{align}
where $\mathcal X$ denotes an arbitrary function of $\mu$. It should be emphasized that this choice of $\mathcal R$ in Eq.~\eqref{eq:Reff} is still a function of $s_1$ and $M_{\rm eff}$ by taking into account Eq.~\eqref{eq:mus1Meff}. Moreover, we allow $M_{\rm eff}$ to take any general form satisfying Eq.~(3.6), and the analysis will proceed accordingly.

For the EM field coupling, the covariance of $g_{\rho\sigma}^{(\mu)}$ must be preserved. This requires modifying the classical EM Hamiltonian into an effective form, $H_{\rm eff}^{EM}$, to ensure that the constraint algebra \eqref{eq:constraint-algebra-GR-eff} and the covariance conditions (i) and (ii) remain valid for
\begin{equation}
 H_{\rm eff} = H_{\rm eff}^{GR} + H_{\rm eff}^{EM}.
\end{equation}
To achieve this, we introduce the ansatz\footnote{Here, this ansatz is introduced by modifying the classical expression \eqref{eq:HEM} by a factor involving $\mu$. In particular, $\sqrt{\mu}$ is introduced in the second term to ensure that the Poisson bracket between $H_{\rm eff}$ and the Gauss constraint $\mathcal G$ retains the same structure as in the classical case, under the replacement $E^1\to \mu E^1$ [see Eq.~\eqref{eq:GH}]. The arbitrary function $\mathcal Y$ can be introduced because it will not change the structure of the constraint algebra (see Appendix \ref{appendix:B}).}
\begin{equation}\label{eq:hemeff}
 H^{EM}_{\rm eff}=\mathcal Y(\mu)\frac{E^2}{2(E^1)^{3/2}}P_\Gamma^2-\frac{\sqrt{\mu E^1}}{E^2}\Gamma \partial_x P_\Gamma,
\end{equation}
with an arbitrary function $\mathcal Y(\mu)$. As shown in Appendix \ref{appendix:B}, the algebra generated by the total constraints reads
\begin{subequations}\label{eq:effectiveconstraintalge}
\begin{align}
 \{\mathcal G[\beta_1],\mathcal G[\beta_2]\}&=0,\label{eq:GG}\\
 \{\mathcal G[\beta],H_x[N^x]\}&=-\mathcal G[N^x\partial_x\beta],\label{eq:GHx}\\
 \{\mathcal G[\beta],H_{\rm eff}[N]\}&=-\mathcal G\left[N\frac{\sqrt{\mu E^1}}{E^2}\partial_x\beta\right],\label{eq:GH}\\
 \{H_x[N^x_1],H_x[N^x_2]\}&=H_x[N^x_1\partial_xN^x_2-N_2^x\partial_xN_1^x],\label{eq:HxHx-eff}\\
 \{H_x[N^x],H_{\rm eff}[N]\}&=H_{\rm eff}[N^x\partial_xN]\label{eq:HxH-eff},\\
 \{H_{\rm eff}[N_1],H_{\rm eff}[N_2]\}&=H_x[\mu S(N_1\partial_xN_2-N_2\partial_xN_1)]\label{eq:HH-eff}.
\end{align}
\end{subequations}
Here, it should be stressed that this algebra holds for all $H_{\rm eff}^{GR}$ and $H^{EM}_{\rm eff}$ with arbitrary selections of the free functions $\mathcal X$, $\mathcal Y$, and $M_{\rm eff}$ satisfying Eq.~\eqref{eq:covariance-mu}, according to the calculations given in Appendix \ref{appendix:B}. According to Eq.~\eqref{eq:effectiveconstraintalge}, the constraints $H_x$ and $H_{\rm eff}$ form a subalgebra as proposed by Eq.~\eqref{eq:constraint-algebra-GR-eff}. Now let us show that the covariance conditions (i) and (ii) are satisfied for $H_{\rm eff}$. At first, the derivatives of $K_1$ are not contained in $H_{\rm eff}$ as that for $H_{\rm eff}^{EM}$, ensuring that the condition (i) is satisfied. Moreover, a direct calculation shows that
\begin{align}\label{eq:covraint-conditon-1}
 \left\{\mu,H_{\rm eff}^{EM}[\alpha N]\right\}=\alpha\left\{\mu,H_{\rm eff}^{EM}[N]\right\}.
\end{align}
Since the derivatives of $K_I$ are not included in $H_{\rm eff}^{EM}$, Eq.~\eqref{eq:covraint-conditon-1} implies
\begin{equation}
 \left\{\mu S,H_{\rm eff}^{EM}[\alpha N]\right\}=\alpha\left\{\mu S,H_{\rm eff}^{EM}[N]\right\}.
\end{equation}
As this relation must hold for $H_{\rm eff}^{GR}$ due to the covariance of the vacuum model, the condition (ii) is thus satisfied.

Now let us turn to the covariance of the EM field in the effective case. Due to the factor $\mu$ in the constraint algebra \eqref{eq:effectiveconstraintalge}, it is easy to show following the procedure given in Sec.~\ref{sec-II-B} that the vector potential $A_\rho$ is no longer covariant, i.e., Eq.~\eqref{eq:covariant-A-class} does not hold any more. To fix the problem, as done for the metric, we will introduce the effective vector potential, denoted by $A_{\rho}^{(\mu)}$, to restore covariance of the EM field. In other words, we will define $A_{\rho}^{(\mu)}$ such that
\begin{equation}\label{eq:covariant-A-eff}
 \delta A_{\rho}^{(\mu)}=\mathcal L_{\alpha \mathfrak N+\beta}A_{\rho}^{(\mu)}+\partial_{\rho}\lambda.
\end{equation}
To get the correct expression of $A_{\rho}^{(\mu)}$, we need to repeat the derivation shown in Sec.~\ref{sec-II-B}, but with the new constraint algebra \eqref{eq:effectiveconstraintalge}. In this case, $\delta \Gamma$ in Eq.~\eqref{eq:deltaEI} and variation of the Lagrangian multipliers in Eq.~\eqref{eq:dNNxEM} become
\begin{equation}
 \begin{aligned}
 \delta\Gamma=&\mathcal L_{\alpha \mathfrak N+\beta}\Gamma+\left(N\frac{\sqrt{\mu E^1}}{E^2}\Gamma+\Psi\right)\partial_x\alpha-\Gamma\mathfrak N^{\rho}\partial_{\rho}\alpha+\partial_x\lambda,\\
 \delta N=&\mathcal L_{\alpha\mathfrak N+\beta}N+N\mathfrak N^{\rho}\partial_{\rho}\alpha,\\
 \delta N^x=&-N^2\mu S \partial_x\alpha-(\mathcal L_{\beta}\mathfrak N)^x,\\
 \delta\Psi=&-N\sqrt{\mu S}(\Psi\partial_x\alpha+\partial_x\lambda)+\beta^x\partial_x\Psi+\mathcal L_{\mathfrak N}(\alpha\Psi+\lambda).\\
 \end{aligned}
\end{equation}
Moreover, $\delta E^I$ in Eq.~\eqref{eq:deltaEI} remains unchanged. Since Eq.~\eqref{eq:covraint-conditon-1} also holds for $H_{\rm eff}$, we have
\begin{equation}
 \begin{aligned}
 \delta\mu=&\{\mu,H_{\rm eff}[\alpha N]+H_x[\beta^x]+\mathcal G[\alpha\Psi+\lambda]\}\\
 =&\alpha\mathcal L_{\mathfrak N}\mu+\mathcal L_{\beta}\mu\\
 =&\mathcal L_{\alpha\mathfrak N+\beta}\mu,
 \end{aligned}
\end{equation}
where we have used the EOM~\eqref{eq:EOMlie} with $H$ replaced by $H_{\rm eff}$ in the second step and the fact that $\mu$ is a spacetime scalar in the third step. Combining all these results, we obtain an explicit form of $A_\rho^{(\mu)}$ as
\begin{equation}\label{eq:Amu-exp}
 A_{\rho}^{(\mu)}\dd x^{\rho} =\left[\Psi+\left(N\frac{\sqrt{\mu E^1}}{E^2}+N^x\right)\Gamma\right]\dd t+\Gamma\dd x,
\end{equation}
such that Eq.~\eqref{eq:covariant-A-eff} is satisfied.

\subsection{Effective dynamics}\label{sec-III-B}

Let us follow the procedure introduced in Secs.~\ref{sec-II-C} and \ref{sec-II-D} to solve the dynamics of the effective model, determined by $H_{\rm eff}^{GR}$ in Eq.~\eqref{eq:covariance-Heff} with $\mathcal R$ given by \eqref{eq:Reff} and $H_{\rm eff}^{EM}$ in Eq.~\eqref{eq:hemeff}. In this case of $\mathcal R\neq0$, $M_{\rm eff}$ is no longer a Dirac observable, while it is in the case of $\mathcal R=0$ for the vacuum model without a cosmological constant considered in~\cite{Zhang:2024khj,Zhang:2024ney}.

Similar as in Sec.~\ref{sec-II-C}, the Gauss constraint $\mathcal G=0$ implies
\begin{equation}
 P_\Gamma=Q.
\end{equation}
Then, choosing the condition $\Gamma=0$ to fix the gauge transformation generated by the Gauss constraint, we get the EOM for $\Psi$ similar to Eq.~\eqref{eq:tildePhi} as
\begin{equation}\label{eq:EOMPsi-1}
 0=\partial_x\Psi+\mathcal Y(\mu)\frac{N E^2}{(E^1)^{3/2}}Q.
\end{equation}
With this result, the effective Hamiltonian constraint and the diffeomorphism constraint become
\begin{equation}\label{eq:barh}
 \begin{aligned}
 \bar H_{\rm eff}=&H_{\rm eff}^{GR}+H_{\rm eff}^Q, \\
 \bar H_x=&H_x^{GR},
 \end{aligned}
\end{equation}
where we introduced
\begin{equation}\label{eq:HQ}
 H_{\rm eff}^Q=H_{\rm eff}^{EM}\Big|_{P_\Gamma=Q}=\mathcal Y(\mu)\frac{E^2 Q^2}{2(E^1)^{3/2}}.
\end{equation}
The reduced diffeomorphism constraint $\bar H_x$ is solved by imposing the areal gauge condition \eqref{eq:arealgauge}, leading once again to the solution given in Eq.~\eqref{eq:solvediff}. Then, as in Sec.~\ref{sec-II-D}, we are allowed to find the stable solution satisfying the stationary condition
\begin{equation}\label{eq:stationaryeff}
 \begin{aligned}
 0&=\dot{E^1}=\left\{E^1,\bar{H}_{\rm eff}[N]+\bar{H}_x[N^x]\right\},\\
 0&=\dot{E^2}=\left\{E^2,\bar{H}_{\rm eff}[N]+\bar{H}_x[N^x]\right\}.
 \end{aligned}
\end{equation}
Under the conditions \eqref{eq:arealgauge} and \eqref{eq:solvediff}, substituting Eq.~\eqref{eq:barh} into Eq.~\eqref{eq:stationaryeff}, we have
\begin{equation}\label{eq:stationaryeff2}
 \begin{aligned}
 0=&G N M_{{\rm eff},s_2}+xN^x,\\
 0=&\partial_x\left(N^xE^2\right)+\frac{GN E^2}{x}\partial_x M_{{\rm eff},s_2}\\
 &-\frac{N E^2 M_{{\rm eff},s_2}}{2x^3}\left(x^4\Lambda\mathcal X_M+G Q^2\mathcal Y_M\right),
 \end{aligned}
\end{equation}
with the abbreviations
\begin{align}
 M_{{\rm eff},s_2}\equiv\partial_{s_2}M_{\rm eff},\quad \mathcal X_M\equiv\frac{\dd \mathcal X}{\dd\mu}\frac{\partial \mu}{\partial M_{\rm eff}},\quad \mathcal Y_M\equiv\frac{\dd \mathcal Y}{\dd\mu}\frac{\partial \mu}{\partial M_{\rm eff}}.
\end{align}
Here, it is used that $\mu$ is a function of $s_1$ and $M_{\rm eff}$ as a consequence of Eq.~\eqref{eq:covarianceequationgeneral2}. Equation~\eqref{eq:stationaryeff2} can be further simplified as
\begin{subequations}\label{eq:stationaryeff3}
\begin{align}
 N^x&=-\frac{GM_{{\rm eff},s_2}}{x}N\label{eq:stationaryeff31},\\
 N&=\frac{x}{E^2}\exp\left[-\int\frac{x^4\Lambda\mathcal X_M(x)+ G Q^2 \mathcal Y_M(x)}{2 G x^2}\dd x\right]\label{eq:stationaryeff32}.
\end{align}
\end{subequations}
It should be emphasized that the functions $M_{{\rm eff},s_2}$, $\mathcal X_M$, and $\mathcal Y_M$ in Eqs.~\eqref{eq:stationaryeff2} and \eqref{eq:stationaryeff3} need to be evaluated under the conditions \eqref{eq:arealgauge} and \eqref{eq:solvediff}.

Now let us consider the effective Hamiltonian constraint. In the conditions \eqref{eq:arealgauge} and \eqref{eq:solvediff}, $\bar H_{\rm eff}=0$ can be rewritten as
\begin{equation}\label{eq:solHeff}
 \begin{aligned}
 \bar H_{\rm eff}=-\frac{E^2}{x}\left[\partial_xM_{\rm eff}-\frac{x^4\Lambda\mathcal X(\mu)+ G Q^2 \mathcal Y(\mu)}{2 G x^2}\right]=0.
 \end{aligned}
\end{equation}
Under the conditions \eqref{eq:arealgauge} and \eqref{eq:solvediff}, $M_{\rm eff}$, $\mathcal X(\mu)$, and $\mathcal Y(\mu)$ become functions of $E^2$ and $K_2$. In other words, Eq.~\eqref{eq:solHeff} establishes a relation between $K_2$ and $E^2$. To solve this equation, a new gauge, such as the Schwarzschild gauge \eqref{eq:SchGauge} or the PG gauge \eqref{eq:PG-gauge}, is needed. Once such a gauge is imposed, Eq.~\eqref{eq:solHeff} can be solved, and $N$ and $N^x$ can be determined using Eq.~\eqref{eq:stationaryeff3}. With $\mu$ obtained from these solutions, the effective metric \eqref{eq:lineelem-eff} then can be constructed.

To get the solution for the effective vector potential $A_{\rho}^{(\mu)}$, we need to solve Eq.~\eqref{eq:EOMPsi-1}. Under the areal gauge, using Eq.~\eqref{eq:stationaryeff32}, Eq.~\eqref{eq:EOMPsi-1} reduces to
\begin{align}\label{eq:EOMPsi}
 \partial_x\Psi+\frac{Q}{x^2}\exp\left[-\int\frac{x^4\Lambda\mathcal X_M(x)+ G Q^2\mathcal Y_M(x)}{2 G x^2}\dd x\right]\mathcal Y(\mu)=0.
\end{align}
Solving this equation to get $\Psi_{\rm eff}$, we finally have
\begin{equation}\label{eq:vectorpotential-eff}
 A_{\rho}^{(\mu)}\dd x^\rho=\Psi_{\rm eff}(x)\dd t,
\end{equation}
where we have chosen $\Gamma=0$.

In principle, the results depend on the choices of $M_{\rm eff}$ and $\mathcal X(\mu)$ and $\mathcal Y(\mu)$. In~\cite{Zhang:2024khj,Zhang:2024ney}, three solutions $M_{\rm eff}$ to Eq.~\eqref{eq:covariance-mu} are proposed, with incorporating polymerization for the classical expression of $M_{\rm eff}$. For convenience, we denote these three solutions by subscripts I, II, and III, respectively. Namely, these three solutions will be denoted by $M_{\rm eff,I}$, $M_{\rm eff,II}$, and $M_{\rm eff,III}$. In what follows, let us consider the resulting effective metrics and vector potentials.

\subsection{Solutions to effective dynamics from $M_{\rm eff,I}$}\label{sec-III-C}

Specifically, the expression for $M_{\rm eff,I}$ is~\cite{Zhang:2024khj,Zhang:2024ney}
\begin{align}\label{eq:MGReffI}
 M_{\rm eff,I}=&\frac{\sqrt{s_1}}{2G}\left[1+\frac{s_1}{\zeta^2}\sin^2\left(\frac{\zeta s_2}{\sqrt{s_1}}\right)-\frac{(s_4)^2}{4}\exp\left(\frac{\mathrm{i}2\zeta s_2}{\sqrt{s_1}}\right)\right]\notag\\
 =&\frac{\sqrt{E^1}}{2G}+\frac{(E^1)^{3/2}\sin^2\left(\frac{\zeta K_2}{\sqrt{E^1}}\right)}{2G\zeta^2}\notag\\
 &-\frac{\sqrt{E^1}}{8G}\left(\frac{\partial_xE^1}{E^2}\right)^2\exp\left(\frac{\mathrm{i}2\zeta K_2}{\sqrt{E^1}}\right).
\end{align}
Here $\zeta$ denotes a quantum parameter. Putting Eq.~\eqref{eq:MGReffI} into the covariance equations \eqref{eq:covariance-mu}, we obtain the corresponding $\mu$ as
\begin{align}
 \mu_{\rm I}=1.
\end{align}
As a consequence, $\mathcal X(\mu_{\rm I})$ and $\mathcal Y(\mu_{\rm I})$ are two constants. Without loss of generality, we set $\mathcal X(\mu_{\rm I})=\mathcal Y(\mu_{\rm I})=1$, since any other nonvanishing values of $\mathcal X(\mu_{\rm I})$ and $\mathcal Y(\mu_{\rm I})$ would simply correspond to redefinitions of $\Lambda$ and $Q$.

Inserting $\mathcal X(\mu)=\mathcal Y(\mu_{\rm I})=1$ into Eq.~\eqref{eq:solHeff}, we get
\begin{equation}\label{eq:HeffasMI}
 \begin{aligned}
 M_{\rm eff,I}=M-\frac{Q^2}{2x}+\frac{\Lambda x^3}{6G},
 \end{aligned}
\end{equation}
with $M$ being the integration constant. Due to the fact that $\mathcal X(\mu_{\rm I})=\mathcal Y(\mu_{\rm I})=1$ leads to $\mathcal X_M=\mathcal Y_M=0$, Eq.~\eqref{eq:stationaryeff32} gives
\begin{equation}\label{eq:prefergauge}
 N=\frac{x}{E^2}.
\end{equation}
Then, Eq.~\eqref{eq:EOMPsi} can be solved easily to get
\begin{equation}
 \Psi_{\rm eff,I}(x)=\frac{Q}{x}.
\end{equation}
Therefore, the effective vector potential \eqref{eq:vectorpotential-eff} reads
\begin{equation}\label{eq:A-eff-I}
 A_\rho^{(\mu_{\rm I})}\dd x^\rho=\frac{Q}{x}\dd t.
\end{equation}

Now, we need to impose an additional gauge fixing condition. Let us choose the Schwarzschild gauge \eqref{eq:SchGauge}, which together with Eq.~\eqref{eq:stationaryeff31} implies
\begin{equation}\label{eq:SchK2E2I}
 -x^2\left[\frac{{\rm i}\zeta e^{\frac{{\rm i}2\zeta K_2}{x}}}{(E^2)^2}-\frac{\sin\left(\frac{2\zeta K_2}{x}\right)}{2\zeta}\right]=0.
\end{equation}
Combining Eq.~\eqref{eq:SchK2E2I} with Eq.~\eqref{eq:MGReffI}, one can eliminate $K_2$ to get
\begin{align}\label{eq:E2SolI}
 E^2&=\pm\frac{x}{\sqrt{\left(1-\frac{2GM_{\rm eff,I}}{x}\right)\left[1+\frac{\zeta^2}{x^2}\left(1-\frac{2GM_{\rm eff,I}}{x}\right)\right]}}.
\end{align}
Plugging Eq.~\eqref{eq:E2SolI} into Eq.~\eqref{eq:prefergauge}, we obtain
\begin{align}
 N^2&=\left(1-\frac{2GM_{\rm eff,I}}{x}\right)\left[1+\frac{\zeta^2}{x^2}\left(1-\frac{2GM_{\rm eff,I}}{x}\right)\right].
\end{align}
Hence, in Schwarzschild coordinates, the effective line element \eqref{eq:lineelem-eff} takes the form
\begin{align}\label{eq:efflineelemI-Sch}
 \dd s^2_{\rm eff,I}&=-f_{\rm eff, I}(x)\dd t^2+f_{\rm eff, I}(x)^{-1}\dd x^2+x^2\dd \Omega^2,
\end{align}
where
\begin{align}
 f_{\rm eff, I}(x)&=\left(1-\frac{2GM_{\rm eff,I}}{x}\right)\left[1+\frac{\zeta^2}{x^2}\left(1-\frac{2GM_{\rm eff,I}}{x}\right)\right]\notag\\
 &=f_{\rm cl}(x)\left[1+\frac{\zeta^2}{x^2}f_{\rm cl}(x)\right]\label{eq:feffIfcl}.
\end{align}
The solution \eqref{eq:efflineelemI-Sch} reduces to the one obtained in~\cite{Lin:2024beb} as $Q\rightarrow0$ and to the vacuum one in~\cite{Zhang:2024khj,Zhang:2024ney} as $Q=\Lambda\rightarrow0$. We can also choose the PG gauge \eqref{eq:PG-gauge} to get the metric. In this gauge, the lapse function becomes $N=1$, according to Eq.~\eqref{eq:prefergauge}. Combining Eq.~\eqref{eq:stationaryeff31} with Eq.~\eqref{eq:HeffasMI} yields
\begin{align}
 N^x=\pm\sqrt{1-f_{\rm eff,I}(x)}.
\end{align}
Thus, in PG coordinates, the effective line element \eqref{eq:lineelem-eff} is
\begin{align}\label{eq:efflineelemI-PG}
 \dd s^2_{\rm eff,I}&=-\dd \tau^2+\left[\dd x\pm\sqrt{1-f_{\rm eff, I}(x)}\,\dd \tau\right]^2+x^2\dd \Omega^2.
\end{align}
It is easy to see that these two solutions \eqref{eq:efflineelemI-Sch} and \eqref{eq:efflineelemI-PG} are related by the coordinate transformation
\begin{align}
 \dd t=\dd \tau\mp\frac{\sqrt{1-f_{\rm eff, I}(x)}}{f_{\rm eff, I}(x)}\,\dd x.
\end{align}
Then, the vector potential \eqref{eq:vectorpotential-eff} in these two coordinates can be related by
\begin{align}
 A_\rho^{(\mu_{\rm I})}\dd x^\rho&=\frac{Q}{x}\dd t
 =\frac{Q}{x}\dd \tau\mp\dd Y_{\rm eff, I}(x),
\end{align}
where $Y_{\rm eff, I}$ is the solution to the ordinary differential equation
\begin{equation}
 \frac{\dd Y_{\rm eff, I}}{\dd x}=\frac{Q}{x}\frac{\sqrt{1-f_{\rm eff, I}(x)}}{f_{\rm eff, I}(x)}.
\end{equation}
Hence, these two vector potentials obtained in these two different gauges are the same up to the gauge term $\dd Y_{\rm eff, I}(x)$, reflecting the covariance of the EM field.

\subsubsection{Properties of spacetime $\dd s^2_{\rm eff,I}$ with $\Lambda=0$}\label{sec:propertyI}

To gain a preliminary insight into the structure of the spacetime, we assume $\Lambda=0$ for simplicity. In addition, we will focus on the physically interesting case where $GM\gg \zeta$, $Q\ll \sqrt{G}M$, corresponding to a black hole with a large mass and a small charge. In this case, $f_{\rm cl}$ will have two distinct roots $x_-<x_+$, respectively, corresponding to the inner horizon $x_-$ and the outer horizon $x_+$ in the classical theory.

\begin{figure}
 \centering
 \includegraphics[width=0.45\textwidth]{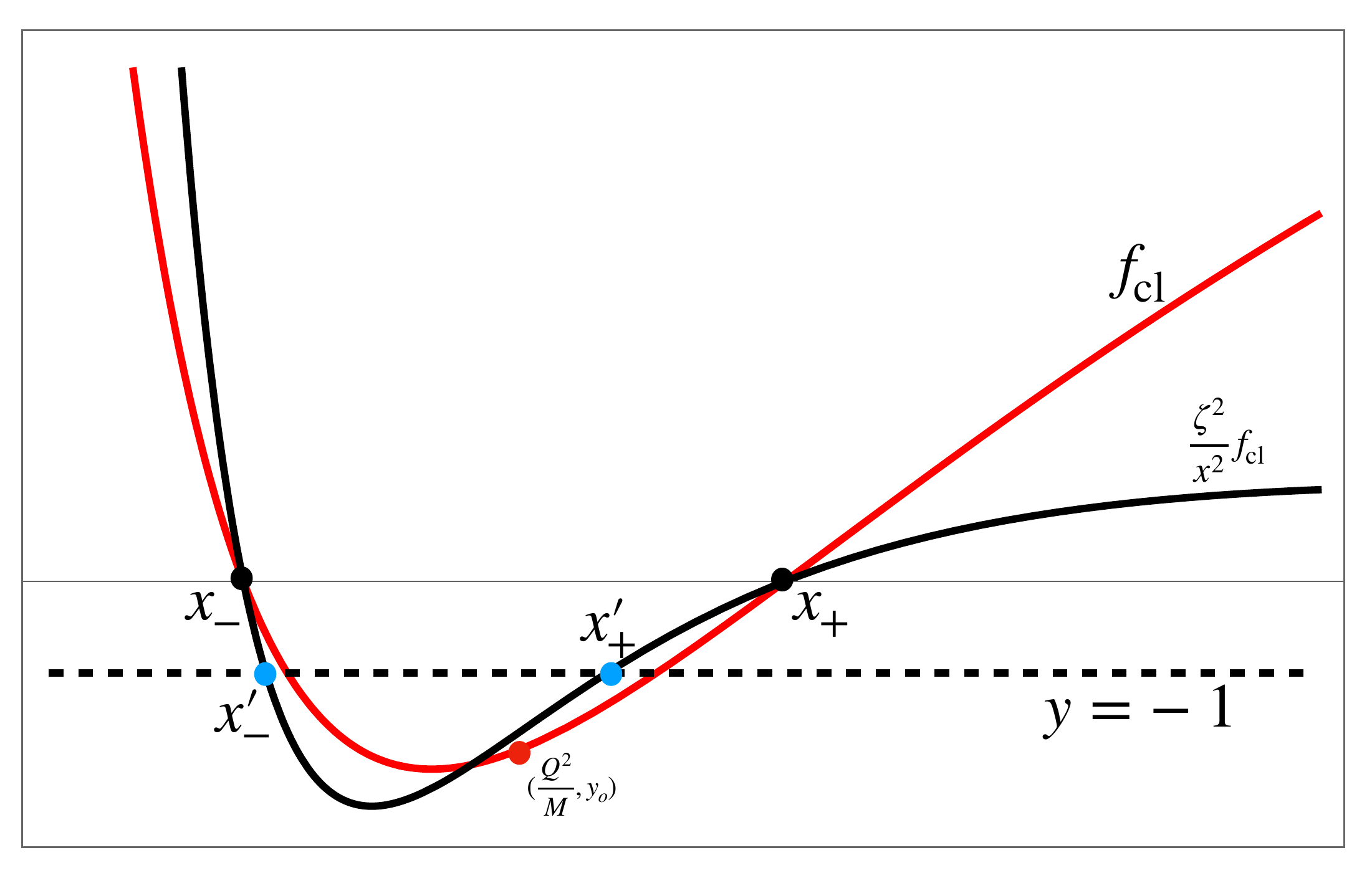}
 \caption{The graph of $f_{\rm cl}$ (red solid line), $\frac{\zeta^2}{x^2}f_{\rm cl}$ (black solid line), and the horizonal line $y=-1$ (black dashed line). According to the facts (1) and (3) given in Sec.~\ref{sec:propertyI}, the graphs pass by the points $(x_\pm, 0)$ (black dot) and $(Q^2/M,y_o)$ (red dot) with $y_o<-1$. In addition, no unexpected features such as oscillations appear in $\frac{\zeta^2}{x^2} f_{\rm cl}$, since it is simply rescaled from $f_{\rm cl}$, which varies smoothly with a single minimum.
 }\label{fig:graph}
\end{figure}

Now let us focus on the horizons of the effective spacetime $\dd s^2_{\rm eff,I}$. According to Eq.~\eqref{eq:feffIfcl}, $f_{\rm eff,I}$ can be factorized such that $f_{\rm cl}$ appears as one of its factors. It then follows that $x_\pm$ remain roots of $f_{\rm eff,I}$. Namely, $x=x_\pm$ still correspond to horizons in the effective spacetime described by $\dd s^2_{\rm eff,I}$. For the second factor $1 + \frac{\zeta^2}{x^2} f_{\rm cl}(x)$, we note that its roots are the intersection of the graph of $y=\frac{\zeta^2}{x^2} f_{\rm cl}(x)$ with the horizontal line $y=-1$. The qualitative behavior of $\frac{\zeta^2}{x^2} f_{\rm cl}(x)$ can be understood from the following facts:
\begin{itemize}
\item[(1)] It shares the same roots $x_\pm$ as $f_{\rm cl}(x)$.
\item[(2)] For $x >x_+$ and $x<x_-$, the function is greater than $0$, and as$x \to \infty$, the function approaches $0$.
\item[(3)] The value of the function at $(x_++x_-)/2 = Q^2/M $ is less than $-1$ in the regime under consideration.
\end{itemize}
These facts, together with the expression of $f_{\rm cl}$, allow us to determine the graph of $\frac{\zeta^2}{x^2} f_{\rm cl}$. As shown in Fig.~\ref{fig:graph}, the function $1 + \frac{\zeta^2}{x^2} f_{\rm cl}(x)$ has two distinct roots, denoted by $x'_\pm$, satisfying $x_- < x'_- < x'_+ < x_+$. As a consequence, the effective spacetime $\dd s^2_{\rm eff,I}$ possesses two additional horizons located at $x = x'_\pm$, which lie between the classical inner and outer horizons. A further analysis of the spacetime structure involves addressing the nontrivial problem of constructing a Penrose diagram for a spacetime with more than one inner horizon~\cite{Schindler:2018wbx}. This issue will be explored in future work.

\subsection{Solutions to effective dynamics from $M_{\rm eff,II}$}\label{sec-III-D}

The second solution $M_{\rm eff,II}$ to Eq.~\eqref{eq:covariance-mu} reads~\cite{Zhang:2024khj,Zhang:2024ney}
\begin{align}\label{eq:MGReffII}
 M_{\rm eff,II}=&\frac{\sqrt{s_1}}{2G}\left[1+\frac{s_1}{\zeta^2}\sin^2\left(\frac{\zeta s_2}{\sqrt{s_1}}\right)-\frac{(s_4)^2}{4}\cos^2\left(\frac{\zeta s_2}{\sqrt{s_1}}\right)\right]\notag\\
 =&\frac{\sqrt{E^1}}{2G}+\frac{(E^1)^{3/2} \sin^2\left(\frac{\zeta K_2}{\sqrt{E^1}}\right)}{2G\zeta^2}\notag\\
 &-\frac{\sqrt{E^1}}{8G}\left(\frac{\partial_xE^1}{E^2}\right)^2\cos^2\left(\frac{\zeta K_2}{\sqrt{E^1}}\right).
\end{align}
The corresponding $\mu$ reads
\begin{align}\label{eq:muII}
 \mu_{\rm II}&=1+\frac{\zeta^2}{E^1}\left(1-\frac{2GM_{\rm eff,II}}{\sqrt{E^1}}\right)\notag\\
 &=1+\frac{\zeta^2}{x^2}\left(1-\frac{2GM_{\rm eff,II}}{x}\right).
\end{align}

To determine the resulting solution, one needs to make a certain gauge fixing. In the Schwarzschild gauge \eqref{eq:SchGauge}, Eq.~\eqref{eq:stationaryeff31} leads to
\begin{equation}\label{eq:SchK2E2II}
 x^2\frac{\zeta^2+(E^2)^2}{2\zeta (E^2)^2}\sin\left(\frac{2\zeta K_2}{x}\right)=0.
\end{equation}
Combining Eqs.~\eqref{eq:SchK2E2II} and \eqref{eq:MGReffII}, we have
\begin{align}
 E^2&=\pm\frac{x}{\sqrt{1-\frac{2GM_{\rm eff,II}}{x}}}.
\end{align}
Using Eq.~\eqref{eq:stationaryeff32}, we get
\begin{align}
 N^2&=\left(\frac{x}{E^2}\exp\left[-\int\frac{x^4\Lambda\mathcal X_M(x)+ G Q^2 \mathcal Y_M(x)}{2 G x^2}\dd x\right]\right)^2\notag\\
 &=\exp\left[-\int\frac{x^4\Lambda\mathcal X_M(x)+ G Q^2 \mathcal Y_M(x)}{G x^2}\dd x\right]f_{\rm eff,II}(x)\notag\\
 &\equiv e^{g_{\rm eff,II}(x)}f_{\rm eff,II}(x),
\end{align}
where we define
\begin{equation*}
 f_{\rm eff,II}(x):=1-\frac{2GM_{\rm eff,II}(x)}{x}.
\end{equation*}
Then, the effective line element \eqref{eq:lineelem-eff} reads
\begin{align}\label{eq:lineelem-Sch-II-eff}
 \dd s^2_{\rm eff, II}&=-e^{g_{\rm eff,II}(x)}f_{\rm eff,II}(x)\dd t^2+\frac{\dd x^2}{\mu_{\rm II}(x) f_{\rm eff,II}(x)}+x^2\dd \Omega^2.
\end{align}
Similarly, in the PG gauge \eqref{eq:PG-gauge}, combining Eqs.~\eqref{eq:stationaryeff31} and \eqref{eq:MGReffII}, we have
\begin{align}
 N^x&=\pm N\sqrt{\mu_{\rm II}\frac{2GM_{\rm eff,II}}{x}}\notag\\
 &=\pm N\sqrt{\mu_{\rm II}(x)\left[1-f_{\rm eff,II}(x)\right]}.
\end{align}
Using Eq.~\eqref{eq:stationaryeff32}, we have
\begin{align}
 N^2&=\left(\frac{x}{E^2}\exp\left[-\int\frac{x^4\Lambda\mathcal X_M(x)+ G Q^2 \mathcal Y_M(x)}{2 G x^2}\dd x\right]\right)^2\notag\\
 &=\exp\left[-\int\frac{x^4\Lambda\mathcal X_M(x)+ G Q^2 \mathcal Y_M(x)}{G x^2}\dd x\right]\notag\\
 &=e^{g_{\rm eff,II}(x)}.
\end{align}
Thus, the effective line element \eqref{eq:lineelem-eff} reduces to
\begin{align}\label{eq:lineelem-PG-II-eff}
 \dd s^2_{\rm eff, II}=&-e^{g_{\rm eff,II}(x)}\dd \tau^2\notag\\
 &+\frac{1}{\mu_{\rm II}(x)}\left(\dd x\pm\sqrt{e^{g_{\rm eff,II}(x)}\mu_{\rm II}(x)\left[1-f_{\rm eff,II}(x)\right]}\dd \tau\right)^2\notag\\
 &+x^2\dd \Omega^2.
\end{align}
These two solutions \eqref{eq:lineelem-Sch-II-eff} and \eqref{eq:lineelem-PG-II-eff} are related by the coordinate transformation
\begin{align}
 \dd t=\dd \tau\mp\frac{\sqrt{\mu_{\rm II}(x)\left[1-f_{\rm eff,II}(x)\right]}}{e^{\frac12g_{\rm eff,II}(x)}\mu_{\rm II}(x)f_{\rm eff,II}(x)}\,\dd x,
\end{align}
reflecting the fact that the effective spacetime is covariant.

On the other hand, Eq.~\eqref{eq:EOMPsi} can be written as
\begin{align}\label{eq:EOMPsi-2}
 \partial_x\Psi_{\rm eff,II}+\frac{Q}{x^2}e^{\frac{1}{2}g_{\rm eff,II}(x)}\mathcal Y(\mu_{\rm II})=0.
\end{align}
Hence, the solution to Eq.~\eqref{eq:EOMPsi-2} can be expressed as
\begin{align}\label{eq:EOMPsi-3}
 \Psi_{\rm eff,II}(x)=-Q\int \frac{e^{\frac{1}{2}g_{\rm eff,II}(x)}}{x^2}\mathcal Y(\mu_{\rm II})\dd x,
\end{align}
where the quantum corrections are incorporated into the expression. Note that the solution $\Psi_{\rm eff,II}(x)$ to Eq.~\eqref{eq:EOMPsi-2} depends only on $x$. Thus
\begin{align}
 A_\rho^{(\mu_{\rm II})}\dd x^\rho&=\Psi_{\rm eff,II}(x)\dd t\notag\\
 &=\Psi_{\rm eff,II}(x)\left[\dd \tau\mp\frac{\sqrt{\mu_{\rm II}(x)\left[1-f_{\rm eff,II}(x)\right]}}{e^{\frac12g_{\rm eff,II}(x)}\mu_{\rm II}(x)f_{\rm eff,II}(x)}\,\dd x\right]\notag\\
 &=\Psi_{\rm eff,II}(x)\dd \tau\mp\dd Y_{\rm eff,II}(x),
\end{align}
where
\begin{align}
 \frac{\dd Y_{\rm eff,II}(x)}{\dd x}=\Psi_{\rm eff,II}(x)\frac{\sqrt{\mu_{\rm II}(x)\left[1-f_{\rm eff,II}(x)\right]}}{e^{\frac12g_{\rm eff,II}(x)}\mu_{\rm II}(x)f_{\rm eff,II}(x)},
\end{align}
indicating the covariance of the EM field.

In the following, we will calculate the values $M_{\rm eff,II}$, $g_{\rm eff,II}(x)$, and $\Psi_{\rm eff,II}(x)$ for different choices of $\mathcal X(\mu_{\rm II})$ and $\mathcal Y(\mu_{\rm II})$, which have exact analytical solutions.

\subsubsection{Case of $\mathcal X(\mu_{\rm II})=\mathcal Y(\mu_{\rm II})=1$}\label{sec:case1}

The simplest case is to choose
\begin{align}\label{eq:XY-1}
 \mathcal X(\mu_{\rm II})=1,\quad\mathcal Y(\mu_{\rm II})=1.
\end{align}
In this case, we have
\begin{equation}\label{eq:XYM-1}
 \mathcal X_M=0,\quad\mathcal Y_M=0.
\end{equation}
Thus, the factor $g_{\rm eff,II}$ takes
\begin{align}
 g_{\rm eff,II}^{(1)}=0.
\end{align}
Equation~\eqref{eq:solHeff} can be solved easily with the solution
\begin{align}
 M_{\rm eff,II}^{(1)}\equiv M_{\rm eff,II}=M-\frac{Q^2}{2x}+\frac{\Lambda x^3}{6G},\label{eq:HeffasMII}
\end{align}
where $M$ denotes the integration constant. Then, Eq.~\eqref{eq:EOMPsi-3} reduces to
\begin{equation}
 \Psi_{\rm eff,II}^{(1)}(x)=\frac{Q}{x}.
\end{equation}

\subsubsection{Case of $\mathcal X(\mu_{\rm II})=1$ and $\mathcal Y(\mu_{\rm II})=\mu_{\rm II}$}\label{sec:caseIi}

It will be interesting to consider a case of $\mathcal X(\mu_{\rm II})=1$ and $\mathcal Y(\mu_{\rm II})\neq 1$. An analytically solvable case is to take
\begin{align}\label{eq:XY-2}
 \mathcal X(\mu_{\rm II})=1,\quad\mathcal Y(\mu_{\rm II})=\mu_{\rm II}.
\end{align}
In this case, we get
\begin{equation}\label{eq:XYM-2}
 \mathcal X_M=0,\quad\mathcal Y_M=-\frac{2\zeta^2G}{x^3}.
\end{equation}
Then, the factor $g_{\rm eff,II}$ reads
\begin{align}
 g_{\rm eff,II}^{(2)}(x)=-\frac{\zeta^2GQ^2}{4 x^4}.
\end{align}
Putting Eq.~\eqref{eq:XY-2} into Eq.~\eqref{eq:solHeff}, we get
\begin{align}\label{eq:solMII}
 M_{\rm eff,II}^{(2)}=&\frac{\Lambda x^3}{6G}+e^{\frac{\zeta^2GQ^2}{4x^4}} \left[\frac{\zeta^2Q^2}{8x^3} E_{\frac{1}{4}}\left(\frac{\zeta^2GQ^2}{4x^4}\right)\right.\notag\\
 &+\frac{Q^2}{8x} \left(1-\frac{\zeta^2\Lambda}{3}\right) E_{\frac{3}{4}}\left(\frac{\zeta^2GQ^2}{4x^4}\right)\notag\\
 &\left.+M-\frac{Q^{3/2}\mathbf{\Gamma}\left(\frac{1}{4}\right)}{4\sqrt{2\zeta}\sqrt[4]{G}}\right],
\end{align}
where $M$ is the integration constant, $E_\alpha(x)=\int_1^\infty e^{-x t}/t^\alpha\dd t$ is the exponential integral function, and $\mathbf{\Gamma}$ denotes the Gamma function. The constant $\frac{Q^{3/2}\Gamma\left(\frac{1}{4}\right)}{4\sqrt{2\zeta}\sqrt[4]{G}}$ is introduced to ensure that $M_{\rm eff,II}^{(2)}$ approaches the correct classical limit, specifically to satisfy
\begin{equation}
 M_{\rm eff,II}^{(2)}=M-\frac{Q^2}{2x}+\frac{\Lambda x^3}{6G}+O\left(\sqrt{\zeta}\right).
\end{equation}
Now we calculate $\Psi_{\rm eff,II}$ from Eq.~\eqref{eq:EOMPsi-3} and obtain
\begin{equation}\label{eq:Psi-eff-2}
 \Psi_{\rm eff,II}^{(2)}=\frac{Q}{x}+\frac{\zeta^2Q \left(9GQ^2-30GM x+20 x^2-20\Lambda x^4\right)}{60x^5}+O\left(\zeta^{5/2}\right).
\end{equation}
According to this result, the quantum gravity effects modify the appearance of the black hole charge in Coulomb’s potential.

\subsubsection{Case of $\mathcal X(\mu_{\rm II})=\mu_{\rm II}$ and $\mathcal Y(\mu_{\rm II})=1$}\label{sec:caseIii}

In the case
\begin{align}\label{eq:XY-3}
 \mathcal X(\mu_{\rm II})=\mu_{\rm II},\quad\mathcal Y(\mu_{\rm II})=1,
\end{align}
we have
\begin{equation}
 \mathcal X_M=-\frac{2\zeta^2G}{x^3},\quad\mathcal Y_M=0.
\end{equation}
Then, the factor $g_{\rm eff,II}$ is
\begin{align}
 g_{\rm eff,II}^{(3)}(x)=\zeta^2\Lambda \ln(x).
\end{align}
The solution to Eq.~\eqref{eq:solHeff} reads
\begin{align}
 M_{\rm eff,II}^{(3)}=&M x^{-\zeta^2 \Lambda}+\frac{Q^2}{2x \left(\zeta^2 \Lambda -1\right)}+\frac{\zeta^2 \Lambda x}{2 \left(\zeta^2G\Lambda +G\right)}\notag\\
 &+\frac{\Lambda x^3}{2 \left(\zeta^2G \Lambda +3G\right)}.
\end{align}
Hence, Eq.~\eqref{eq:EOMPsi-3} leads to
\begin{equation}\label{eq:Psi-eff-3}
 \Psi_{\rm eff,II}^{(3)}=\frac{Q}{x}+\frac{\zeta^2Q\Lambda}{x}[1+\ln(x)]+O(\zeta^3).
\end{equation}

\subsubsection{Properties of spacetime $\dd s_{\rm eff, II}^2$ with $\Lambda=0$}\label{sec:metricIIspa}

In this section, we analyze the properties of the spacetime $\dd s_{\rm eff, II}^2$, using the case presented in Sec.~\ref{sec:case1} as a concrete example. For the case discussed in Secs.~\ref{sec:caseIi} and \ref{sec:caseIii}, the analysis is similar but technically more involved.

For simplicity, we again assume $\Lambda=0$, $GM\gg \zeta$, and $Q\ll \sqrt{G}M$. According to Eq.~\eqref{eq:lineelem-Sch-II-eff}, the metric $\dd s_{\rm eff, II}^2$ under the Schwarzschild-like coordinates takes the form
\begin{equation}\label{eq:metricII}
 \dd s_{\rm eff, II}^2=-f(x)\dd t^2+g(x)^{-1}\dd x^2+x^2\dd\Omega^2,
\end{equation}
with $f\neq g$. In the case we considered, $f$ is proportional to $f_{\rm cl}$ by a nonvanishing factor. Thus, as discussed in Sec.~\ref{sec:propertyI}, it possesses two positive roots: $x_\pm$. Moreover, $g(x)$ is equal to $f_{\rm eff,I}$, and consequently, it owns four positive roots: $x_\pm$ and $x'_\pm$ with $x_-<x_-'<x_+'<x_+$ (see Fig.~\ref{fig:graph}). Although all of the roots give coordinate singularities, only the roots of $f(x)$, namely, $x_\pm$, correspond to horizons. According to the discussion in Sec.~\ref{sec:propertyI}, we get $f(x'_+)<0$, implying that the hypersurface $x=x'_+$ is spacelike. Indeed, one may regard $x=x'_+$ as the limit of a series of hypersurfaces $x=c$ as $x\to x'_+$. In this limit, the induced metric on $x=x'_+$ is given by
\begin{equation}
 \dd s^2\big|_{x=x'_+}=\lim_{c\to x'_+}\dd s^2\big|_{x=c}=-f(x'_+)\dd t^2+(x'_+)^2\dd\Omega^2,
\end{equation}
which clearly confirms its spacelike character.

Before investigating the extension of the spacetime $\dd s_{\rm eff}^2$, it should be stressed that the domain $x > 0$ is divided into five separate regions by the four coordinate singularities $x_\pm$, $x'_\pm$, and the maximal extension of one of these regions does not necessarily include the others. In our analysis, we begin with the region $x > x_+$. Since $x = x_+$ corresponds to a coordinate singularity, extending the spacetime across this surface requires a new coordinate chart that remains regular at $x = x_+$. The PG-like coordinate system defined by Eq.~\eqref{eq:lineelem-PG-II-eff} provides such a coordinate chart. With this coordinate, the spacetime will be extended to include the region $x'_+<x<x_+$. To extend the space beyond $x=x'_+$, we follow the method suggested in~\cite{Alonso-Bardaji:2023niu} for a similar context. Specifically, we introduce a new coordinate $z(x)$ defined via
\begin{equation}\label{eq:dzx}
 \left(\frac{\dd z(x)}{\dd x}\right)^2=f(x)g(x)^{-1}=\mu_{\rm eff,II}(x)^{-1}.
\end{equation}
In terms of the new coordinates $(t, z)$, the metric takes the form
\begin{equation}
 \dd s^2=-f(z)\dd t^2+f(z)^{-1}\dd z^2+x(z)^2\dd\Omega^2,
\end{equation}
where $x(z)$ is the inverse function of $z(x)$ determined in Eq.~\eqref{eq:dzx}. Since $x = x'_+$ is a root of $g(x)$, the function $x(z)$ exhibits a turning point at $x(z) = x'_+$. This implies that the extension beyond $x = x'_+$ does not penetrate into the region with $x < x_+$, but instead reflects back. Since the classical timelike singularity lies within the region $x\in(0,x_-)$ where $x_- < x'_+$, the spacetime in our model avoids this singularity and is thus nonsingular. The Penrose diagram corresponds to that proposed for the second model in our previous work~\cite{Zhang:2024khj}, which resembles the gluing of two Kruskal spacetimes across a black-to-white hole transition surface.

\subsection{Solutions to effective dynamics from $M_{\rm eff,III}$}\label{sec-III-E}

The third solution $M_{\rm eff,III}$ to Eq.~\eqref{eq:covariance-mu} is given by~\cite{Zhang:2024ney}
\begin{equation}\label{eq:MGReffIII}
 \begin{aligned}
 M_{\rm eff,III}=&\frac{(s_1)^{3/2}}{2G\zeta^2}\sin\left[\frac{\zeta^2}{s_1}\left(1+(s_2)^2-\frac{(s_4)^2}{4}\right)\right]\\
 =&\frac{(E^1)^{3/2}}{2G\zeta^2}\sin\left[\frac{\zeta^2}{E^1}\left(1+(K_2)^2-\frac{(\partial_xE^1)^2}{4(E^2)^2}\right)\right].
 \end{aligned}
\end{equation}
The corresponding $\mu$ is
\begin{align}\label{eq:muIII}
 \mu_{\rm III}&=1-\frac{4\zeta^4(GM_{\rm eff,III})^2}{(s_1)^3}=1-\frac{4\zeta^4(GM_{\rm eff,III})^2}{x^6}.
\end{align}

In the Schwarzschild gauge \eqref{eq:SchGauge}, Eq.~\eqref{eq:stationaryeff31} yields
\begin{equation}\label{eq:SchK2E2III}
 x\cos\left[\frac{\zeta^2}{x^2}\left(1-\frac{x^2}{(E^2)^2}+(K_2)^2\right)\right]K_2=0.
\end{equation}
Combining Eqs.~\eqref{eq:SchK2E2III} and \eqref{eq:MGReffIII}, we obtain
\begin{align}
 E^2&=\pm\frac{x}{\sqrt{1-\frac{x^2}{\zeta^2}\left[n\pi+(-1)^n\arcsin\left(\frac{2\zeta^2GM_{\rm eff,III}}{x^3}\right)\right]}}
\end{align}
Using Eq.~\eqref{eq:stationaryeff32}, we have
\begin{align}
 N^2&=\exp\left[-\int\frac{x^4\Lambda\mathcal X_M(x)+ G Q^2 \mathcal Y_M(x)}{G x^2}\dd x\right]f_{\rm eff,III}^{(n)}(x)\notag\\
 &\equiv e^{g_{\rm eff,III}(x)}f_{\rm eff,III}^{(n)}(x),
\end{align}
where we introduced $f_{\rm eff,III}^{(n)}(x)$ as
\begin{equation}\label{eq:fIII}
 f_{\rm eff,III}^{(n)}(x)=1-\frac{x^2}{\zeta^2}\left[n\pi+(-1)^n\arcsin\left(\frac{2\zeta^2GM_{\rm eff,III}}{x^3}\right)\right].
\end{equation}
Then, the effective line element \eqref{eq:lineelem-eff} reads
\begin{align}\label{eq:lineelem-Sch-III-eff}
 \dd s^2_{\rm eff,III}&=-e^{g_{\rm eff,III}(x)}f_{\rm eff,III}^{(n)}(x)\dd t^2+\frac{\dd x^2}{\mu_{\rm III}(x) f_{\rm eff,III}^{(n)}(x)}+x^2\dd \Omega^2.
\end{align}
Similarly, in the PG gauge \eqref{eq:PG-gauge}, combining Eqs.~\eqref{eq:stationaryeff31} and \eqref{eq:MGReffIII}, we obtain
\begin{align}
 N^x&=\pm N\sqrt{\mu_{\rm III}(x)\left[1-f_{\rm eff,III}(x)\right]}.
\end{align}
Using Eq.~\eqref{eq:stationaryeff32}, we have
\begin{align}
 N^2&=\exp\left[-\int\frac{x^4\Lambda\mathcal X_M(x)+ G Q^2 \mathcal Y_M(x)}{G x^2}\dd x\right]\notag\\
 &=e^{g_{\rm eff,III}(x)}.
\end{align}
Thus, the effective line element \eqref{eq:lineelem-eff} reduces to
\begin{align}\label{eq:lineelem-PG-III-eff}
 \dd s^2_{\rm eff,III}=&-e^{g_{\rm eff,III}(x)}\dd \tau^2\notag\\
 &+\frac{1}{\mu_{\rm III}(x)}\left(\dd x\pm\sqrt{e^{g_{\rm eff,III}(x)}\mu_{\rm III}(x)\left[1-f_{\rm eff,III}^{(n)}(x)\right]}\dd \tau\right)^2\notag\\
 &+x^2\dd \Omega^2.
\end{align}
The remaining calculations are the same as those in Sec.~\ref{sec-III-D}, just by replacing subscript II by III.

\subsubsection{Properties of spacetime $\dd s^2_{\rm eff,III}$ with $\Lambda=0$}\label{sec:propertyIII}

For concrete analysis, we will consider the case with $\mathcal X=1=\mathcal Y$, leading to $N=1$, and
\begin{equation}
 M_{\rm eff,III}=M-\frac{Q^2}{2x}+\frac{\Lambda x^3}{6G}.
\end{equation}
Let us still focus on the case where $\Lambda=0$, $\sqrt{G}Q\ll GM$, and $GM\gg \zeta^2$.

For the metric $\dd s_{\rm eff,III}^2$ in Schwarzschild-like coordinates, as given in Eq.~\eqref{eq:lineelem-Sch-III-eff}, it retains the same form as Eq.~\eqref{eq:metricII}. Consequently, the procedure for extending the spacetime follows exactly the same steps as those described in Sec.~\ref{sec:metricIIspa}.

To investigate the maximally extended spacetime of $\dd s_{\rm eff,III}^2$, let us start with $n=0$. Due to the presence of the arcsine function in $f_{\rm eff,III}^{(0)}$, $x$ must lie within the domain $[x_{\rm min},\infty)$ where $x_{\rm min}$ is defined by
\begin{equation}
 \frac{2\zeta^2GM_{\rm eff,III}(x_{\rm min})}{x_{\rm min}^3}=1.
\end{equation}
Within this domain we have
\begin{equation}\label{eq:mugg0}
 \mu(x)>0,\ \forall x>x_{\rm min},\quad \text{and}\quad\mu(x_{\rm min})=0.
\end{equation}
In the regime with $\Lambda=0$, $\sqrt{G}Q\ll GM$, and $GM\gg \zeta^2$, $f_{\rm eff,III}^{(0)}$ possesses two roots, denoted by $x_+>x_-$. Due to Eq.~\eqref{eq:mugg0}, the PG coordinates given in \eqref{eq:lineelem-PG-III-eff} are well-defined across $x=x_\pm$. Thus, the extension of the region $x>x_+$ can be achieved by using the PG coordinates, and the resulting extension will cover the entire domain $x>x_{\rm min}$. To extend the spacetime beyond $x=x_{\rm min}$, we again introduce the coordinate $z(x)$ such that
\begin{equation}
 \left(\frac{\dd z}{\dd x}\right)^2=\mu_{\rm III}^{-1}.
\end{equation}
Then, as in Sec.~\ref{sec:metricIIspa}, $x(z)$ exhibit a turning point at $x(z)=x_{\rm min}$. However, unlike the case discussed in Sec.~\ref{sec:metricIIspa}, a subtle issue arises here due to the presence of the arcsine function in $f_{\rm eff,III}^{(n)}$: after the turning point, the line element should transition to the branch of the arcsine function corresponding to $n = 1$. This transition is essential to ensure the smoothness of the metric across the turning point. Indeed, by choosing the $n = 1$ branch after the turning point, the metric, in the $(t,z)$ coordinates, reads
\begin{equation}\label{eq:metricIII}
 \dd s^2_{\rm eff,III}=\left\{\begin{array}{cc}
 -f_{\rm eff,III}^{(0)}(z)\dd t^2+\frac{\dd x^2}{f_{\rm eff,III}^{(0)}(z)}+x(z)^2\dd \Omega^2,& z\geq z_o\\
 -f_{\rm eff,III}^{(1)}(z)\dd t^2+\frac{\dd x^2}{f_{\rm eff,III}^{(1)}(z)}+x(z)^2\dd \Omega^2,& z\leq z_o
 \end{array}
 \right.
\end{equation}
where $z_o$ is the value of $z$ at the turning point and $z>z_o$ corresponds to the region before the turning point. Then, the smoothness of the metric is ensured due to
\begin{equation}
 \lim_{z\to z_o^+}\frac{\dd^n}{\dd z^n}f_{\rm eff,III}^{(0)}=\lim_{z\to z_o^-}\frac{\dd^n}{\dd z^n}f_{\rm eff,III}^{(1)}.
\end{equation}
In addition, following a similar line of reasoning, we find that if the arcsine function is constrained to remain in the $n = 0$ branch after the turning point, the resulting metric becomes nondifferentiable due to
\begin{equation}
 \lim_{z\to z_o^+}\frac{\dd}{\dd z}f_{\rm eff,III}^{(0)}=-\lim_{z\to z_o^-}\frac{\dd}{\dd z}f_{\rm eff,III}^{(0)}.
\end{equation}
This demonstrates the necessity of transitioning to the $n = 1$ branch to ensure the smoothness of the metric. The space is complete after the one corresponding to $n=1$ is included. The resulting Penrose diagram coincides with the one presented in our previous work~\cite{Zhang:2024ney}, resembling that of the Kruskal spacetime but with the central singularity replaced by an asymptotically de Sitter region. This de Sitter region arises from the presence of the $n\pi$ term, which is nonvanishing for $n=1$, in Eq.~\eqref{eq:fIII}.

\section{Covariance of gravity coupled to a general matter field}\label{sec-IV}

In this section, let us discuss the covariance of gravity coupled to a general matter field.

As done for the EM field, to maintain covariance for the effective metric $g_{\rho\sigma}^{(\mu)}$, the total Hamiltonian constraint needs to share the similar algebraic structure as Eq.~\eqref{eq:effectiveconstraintalge}, and to satisfy the covariance conditions presented in Sec.~\ref{sec-III-A}. Let us assume that this can be implemented by modifying the classical matter Hamiltonian into the effective one $H^{\rm matter}_{\rm eff}$, which incorporates the quantity $\mu$. Since the classical matter Hamiltonian usually depends only on $E^1$ and $E^2$ as the components of the spatial metric and since $\mu$ depends only on $E^I$ and $K_2$ according to the covariance equation \eqref{eq:covariance-mu}, incorporating $\mu$ into $H^{\rm matter}_{\rm eff}$ does not introduce a dependence on derivatives of $K_1$. Thus, the covariance condition (i) is satisfied. Moreover, it can be verified that
\begin{align}
 \left\{\mu,\alpha Nf(\mu,E^1,E^2)\right\}=\alpha\left\{\mu,N f(\mu,E^1,E^2)\right\},
\end{align}
for an arbitrary function $f$. As a consequence, we get
\begin{align}\label{eq:covraint-conditon-2}
 \left\{\mu,H_{\rm eff}^{\rm matter}[\alpha N]\right\}=\alpha\left\{\mu,H_{\rm eff}^{\rm matter}[N]\right\}.
\end{align}
Since Eq.~\eqref{eq:covraint-conditon-2} remains valid with replacing $H_{\rm eff}^{\rm matter}$ by $H_{\rm eff}^{GR}$ given in Eq.~\eqref{eq:hemeff}, it is concluded that Eq.~\eqref{eq:covraint-conditon-2} is satisfied by the total effective Hamiltonian constraint $H_{\rm eff}^{GR}+H_{\rm eff}^{\rm matter}$. Namely, the covariance condition (ii) is also satisfied.

According to the above discussion, to maintain the covariance for the effective metric $g_{\rho\sigma}^{(\mu)}$ in the coupling model with a general matter field whose classical Hamiltonian depends only on $E^1$ and $E^2$, it is sufficient to modify the classical matter Hamiltonian by incorporating $\mu$ {and requiring that} the total Hamiltonian constraint retains the similar algebraic structure as Eq.~\eqref{eq:effectiveconstraintalge}. As discussed in Appendix~\ref{appendix:B}, Eqs.~\eqref{eq:HxHx-eff} and \eqref{eq:HxH-eff} follow directly from the fact that the diffeomorphism constraint generates spatial diffeomorphism transformations. Therefore, as long as the classical form of the diffeomorphism constraint is preserved, these two equations hold automatically. Consequently, our attention can be focused on the final equation, Eq.~\eqref{eq:HH-eff}. Based on the derivations in Appendix~\ref{appendix:B}, the cross term
\begin{align}
 \{H_{\rm eff}^{\rm matter}[N_1]+H^{GR}_{\rm eff}[N_1],H_{\rm eff}^{\rm matter}[N_2]+H^{GR}_{\rm eff}[N_2]\}\notag
\end{align}
vanishes since $H_{\rm eff}^{\rm matter}$ depends on $E^I$ and $\mu$. In addition, the bracket $\{H^{\mathrm{GR}}_{\mathrm{eff}}[N_1], H^{\mathrm{GR}}_ {\mathrm{eff}}[N_2]\}$ already takes the correct form as given in Eq.~\eqref{eq:HH-eff}. Therefore, to ensure that the full constraint algebra retains the correct structure, it is sufficient to require that the modified matter Hamiltonian have the correct Poisson bracket structure with itself.

It is important to note that in the coupling model with matter fields, maintaining covariance also requires the matter fields themselves to be covariant. This needs us to define an effective 4D matter field that may depend on phase space variables and the Lagrange multiplier, similar to the treatment of the field $A^{(\mu)}_\rho$. In general, the precise definition of the effective matter field requires further investigation.

\section{Summary and discussion}\label{sec-V}

In this paper, we successfully extended the approach introduced in~\cite{Zhang:2024khj,Zhang:2024ney} for restoring covariance in the effective quantum BH models to the electrovacuum case with a cosmological constant. In Sec.~\ref{sec-II}, we first recalled the Hamiltonian framework of gravity coupled to the EM field in the spherically symmetric case. Specifically, we introduced the EM diffeomorphism constraint \eqref{eq:HxEM} following Refs.~\cite{Louko:1996dw,Tibrewala:2012xb}, such that the total diffeomorphism constraint generates the spatial diffeomorphism transformations for gravity and the EM field. Additionally, the EM Gauss constraint is incorporated into the Hamiltonian \eqref{eq:HEM} of the EM field, ensuring that the total Hamiltonian and diffeomorphism constraints of the coupling model form a subalgebra \eqref{eq:HxHx}--\eqref{eq:HH} as that of the vacuum model in~\cite{Zhang:2024khj,Zhang:2024ney}. As a result, the analysis for spacetime covariance in the coupling model will be parallel to that in the vacuum case. The covariance condition \eqref{eq:covariant-g-class} for gravity and the one \eqref{eq:covariant-A-class} for the EM field have been examined in the classical theory. By solving the Gauss constraint, we derived the solutions \eqref{eq:lineelem-Sch} and \eqref{eq:lineelem-PG} to the EOM in different gauge fixings, corresponding to the RN--(anti--)de Sitter metric.

In Sec.~\ref{sec-III}, we studied covariance of the effective coupling model. To ensure validity of the constraint algebra \eqref{eq:constraint-algebra-GR-eff} and the covariance conditions (i) and (ii), the classical Hamiltonian \eqref{eq:HEM} of the EM field has been modified into the effective one \eqref{eq:hemeff} by incorporating $\mu$, leading to the covariance of the effective metric. To require the effective vector potential satisfying the covariance equation \eqref{eq:covariant-A-eff}, we obtained an explicit form \eqref{eq:Amu-exp} of $A_\rho^{(\mu)}$ by a detailed derivation. Based on the covariance effective coupling model, we further explored its exact solutions. Thanks to the three solutions $M_{\rm eff,I}$, $M_{\rm eff,II}$, and $M_{\rm eff,III}$ to Eq.~\eqref{eq:covariance-mu} for building the effective Hamiltonian constraints of gravity in the vacuum model, the corresponding exact solutions for the effective coupling model have been founded in Sec.~\ref{sec-III-C}, Sec.~\ref{sec-III-D}, and Sec.~\ref{sec-III-E}. In the solutions corresponding to $M_{\rm eff,I}$, the vector potential $A_\rho^{(\mu_{\rm I})}$ remains its classical form \eqref{eq:vectorpotentialSol} in a specific gauge choice, while the effective metric receives the quantum corrections with the line element \eqref{eq:efflineelemI-Sch} in Schwarzschild coordinates and \eqref{eq:efflineelemI-PG} in PG ones. Interestingly, in the solutions corresponding to $M_{\rm eff,II}$ (and $M_{\rm eff,III}$), there exist effective vector potentials that incorporate quantum gravity effects, as shown in Eqs.~\eqref{eq:Psi-eff-2} and \eqref{eq:Psi-eff-3}, due to the existence of arbitrary functions $\mathcal X(\mu)$ and $\mathcal Y(\mu)$. Hence, in these cases, quantum gravity effects could affect the EM field, besides changing the structure of spacetime. Moreover, the influence of quantum gravity effects on the spacetime structure has been preliminarily explored in Secs.~\ref{sec:propertyI}, \ref{sec:metricIIspa}, and \ref{sec:propertyIII}, focusing on the case with a vanishing cosmological constant, a small charge, and a large black hole mass. Although only a preliminary investigation, this analysis raises several open questions, for example, how to construct Penrose diagrams for spacetimes with multiple horizons, and what the implications are of the singularity resolution in $\dd s_{\rm eff,II}^2$ and $\dd s_{\rm eff,III}^2$. A more comprehensive study, including an investigation of the spacetime structure in other regimes with a nonvanishing cosmological constant and a large charge, will be valuable, and we leave these questions for future work.

Finally, how to extend covariance to a general matter field coupling was discussed in Sec.~\ref{sec-IV}. The main result can be summarized as follows: to preserve covariance for the effective metric $g_{\rho\sigma}^{(\mu)}$ in the coupling model with a general matter field whose classical Hamiltonian depends only on $E^1$ and $E^2$, it is sufficient to modify the classical matter Hamiltonian by incorporating $\mu$, ensuring that the total Hamiltonian constraint maintains the similar algebraic structure as Eq.~\eqref{eq:effectiveconstraintalge}.

\begin{acknowledgments}
This work is supported in part by NSFC Grants No. 12165005, No. 12275022, and ``the Fundamental Research Funds for the Central Universities".
\end{acknowledgments}

\appendix

\section{Covariance of the classical theory}\label{app:A}

Observing that the change $\delta g_{\rho\sigma}$ of $g_{\rho\sigma}$ given by Eq.~\eqref{eq:lineelem} includes $\delta E^I$, $\delta N$, and $\delta N^x$, let us first calculate $\delta E^I$ by definition~\eqref{eq:deltaW}. As discussed in~\cite{Zhang:2024khj,Zhang:2024ney}, it is sufficient to consider the case where $\alpha$, $\beta^x$, and $\lambda$ are phase space independent. Then, for $\alpha$, $\beta^x$, and $\lambda$ being phase space dependent, all of the results will still apply, but only on-shell, i.e., when the constraints are satisfied. We have
\begin{align}\label{eq:de111app}
 \delta E^I=\left\{E^I,H[\alpha N]\right\}+\mathcal L_\beta E^I,
\end{align}
where we used that $\{E^I(x),\mathcal G(y)\}=0$ and the fact that $H_x$ generates the spatial diffeomorphism transformation. Since $H$ does not include the derivatives of $K_I$, the first term on the right-hand side of Eq.~\eqref{eq:de111app} reduces to
\begin{equation}
 \left\{E^I(x),H[\alpha N]\right\}=\alpha(x)\left\{E^I,H[N]\right\}=\alpha\mathcal L_{\mathfrak N}E^I,
\end{equation}
where Eq.~\eqref{eq:EOMlie} is used in the last step. Substituting this result into Eq.~\eqref{eq:de111app} and noting that $E^1$ is a scalar but $E^2$ is a scalar density, we finally get
\begin{equation}\label{eq:deltaEIapp}
 \begin{aligned}
 \delta E^1&=\mathcal L_{\alpha\mathfrak N+\beta}E^1,\\
 \delta E^2&=\mathcal L_{\alpha\mathfrak N+\beta}E^2-E^2\mathfrak N^\rho\partial_\rho\alpha.
 \end{aligned}
\end{equation}
To calculate $\delta N$ and $\delta N^x$, we consider that the gauge transformation \eqref{eq:gauge} will map a solution $K_I(t,x)$, $E^I(t,x)$ to a new set of solutions
\begin{equation}\label{eq:EOMEMNNxapp}
 \begin{aligned}
 \tilde K_I=&K_I+\epsilon\left\{K_I,H[\alpha N]+H_x[\beta^x]\right\}+O(\epsilon^2),\\
 \tilde E^I=&E^I+\epsilon\left\{E^I,H[\alpha N]+H_x[\beta^x]\right\}+O(\epsilon^2).\\
 \end{aligned}
\end{equation}

As discussed in Sec.~\ref{sec-II-B}, the transformed solutions must satisfy Hamilton’s equations with respect to the modified lapse function $\tilde{N} = N + \epsilon\, \delta N$ and shift vector $\tilde{N}^x = N^x + \epsilon\, \delta N^x$. It is worth noting that $\delta N$ and $\delta N^x$ were derived in the vacuum case in~\cite{Zhang:2024khj,Zhang:2024ney}. These results, as presented by the first two equations in Eq.~\eqref{eq:dNNxEM}, can be directly adopted here, as they rely on two key ingredients that remain intact in the current nonvacuum model: (a) the fact that the diffeomorphism constraint generates spatial diffeomorphism transformations, which results in EOM for $E^I$ and $K_I$ taking the same forms as Eq.~\eqref{eq:EOMlie}; and (b) the structure of constraint algebra, which is the same as Eqs.~\eqref{eq:HxHx}--\eqref{eq:HH}.

Similar to getting Eq.~\eqref{eq:deltaEIapp}, we have
\begin{equation}\label{eq:deltagammaapp}
 \delta\Gamma=\mathcal L_{\alpha \mathfrak N+\beta}\Gamma+\left(N\frac{\sqrt{E^1}}{E^2}\Gamma+\Psi\right)\partial_x\alpha-\Gamma\mathfrak N^{\rho}\partial_{\rho}\alpha+\partial_x\lambda.
\end{equation}
To calculate $\delta\Psi$, we again consider that the gauge transformation \eqref{eq:gauge} will map a solution $\Gamma(t,x)$, $P_\Gamma(t,x)$ to a new set of solutions
\begin{equation}\label{eq:EOMEMapp}
 \begin{aligned}
 \tilde \Gamma=&\Gamma+\epsilon\delta\Gamma+O(\epsilon^2),\\
 \tilde P_\Gamma=&P_\Gamma+\epsilon\delta P_\Gamma+O(\epsilon^2).
 \end{aligned}
\end{equation}
They need to satisfy Hamilton's equations with respect to a new lapse function $\tilde N=N+\epsilon\delta N$, a new shift vector $\tilde N^x=N^x+\epsilon\delta N^x$, and a new Lagrangian multiplier $\tilde\Psi=\Psi+\epsilon\delta\Psi$, where $\delta N$ and $\delta N^x$ have been given by Eq.~\eqref{eq:dNNxEM}.

Substituting $\tilde \Gamma$ and $\tilde P_\Gamma$ given by Eq.~\eqref{eq:EOMEMapp} into Eq.~\eqref{eq:EOMlie} and comparing the first order results in $\epsilon$ on both sides, we get
\begin{equation}\label{eq:dPG}
 \begin{aligned}
 &\mathcal L_{\delta \mathfrak N}X+\mathcal L_{\mathfrak N}\delta X=\{X,H[\delta N]+\mathcal G[\delta \Psi]\}\\
 &+\Big\{\left\{X,H[N]+\mathcal G[\Psi]\right\},H[\alpha N]+H_x[\beta^x]+\mathcal G[\alpha\Psi+\lambda]\Big\},
 \end{aligned}
\end{equation}
where $X$ is used to represent either $\Gamma$ or $P_\Gamma$. Due to the fact that $\delta\mathfrak N=-\delta N^x\partial_x$, the first term on the left-hand side of Eq.~\eqref{eq:dPG} reduces to
\begin{equation}\label{eq:result1}
 \mathcal L_{\delta \mathfrak N}X=\left\{X,H_x[\delta\mathfrak N^x]\right\}.
\end{equation}
Then, we compute the second term on the left-hand side of Eq.~\eqref{eq:dPG} and get
\begin{equation}\label{eq:dPG-1}
 \begin{aligned}
 &\mathcal L_{\mathfrak N}\delta X\\
 =&\mathcal L_{\mathfrak N}\Big\{X,H[\alpha N]+H_x[\beta^x]+\mathcal G[\alpha\Psi+\lambda]\Big\}\\
 =&\Big\{\left\{X,H[\alpha N]+H_x[\beta^x]+\mathcal G[\alpha\Psi+\lambda]\right\},H[N]+\mathcal G[\Psi]\Big\}\\
 &+\{X,H[\mathcal L_{\mathfrak N}(\alpha N)]+H_x[\mathcal L_{\mathfrak N}\beta^x]+\mathcal G[\mathcal L_{\mathfrak N}(\alpha\Psi+\lambda)]\}.
 \end{aligned}
\end{equation}
Note that in the last step in Eq.~\eqref{eq:dPG-1}, the first term contains the Lie derivatives for the phase-space-dependent parts in $\delta X$, and the second term includes the Lie derivative for the phase-space-independent variables. Applying the Jacobi identity for Poisson brackets to Eq.~\eqref{eq:dPG-1}, we get
\begin{equation}\label{eq:result3}
 \begin{aligned}
 \mathcal L_{\mathfrak N}\delta X=&\Big\{\left\{X,H[N]+\mathcal G[\Psi]\right\},H[\alpha N]+H_x[\beta^x]+\mathcal G[\alpha\Psi+\lambda]\Big\}\\
 &+\Big\{X,\left\{H[\alpha N]+H_x[\beta^x]+\mathcal G[\alpha\Psi+\lambda],H[N]+\mathcal G[\Psi]\right\}\Big\}\\
 &+\{X,H[\mathcal L_{\mathfrak N}(\alpha N)]+H_x[\mathcal L_{\mathfrak N}\beta^x]+\mathcal G[\mathcal L_{\mathfrak N}(\alpha\Psi+\lambda)]\}.
 \end{aligned}
\end{equation}
The second term on the right-hand side of Eq.~\eqref{eq:result3} contains a Poisson bracket between constraints, and thus it can be evaluated by applying the constraint algebra \eqref{eq:GG}--\eqref{eq:HH-eff}. The result reads
\begin{equation}\label{eq:result2}
 \begin{aligned}
 \mathcal L_{\mathfrak N}\delta X=&\Big\{\left\{X,H[N]+\mathcal G[\Psi]\right\},H[\alpha N]+H_x[\beta^x]+\mathcal G[\alpha\Psi+\lambda]\Big\}\\
 &+\left\{X,\mathcal G\left[N \sqrt{S}(-\Psi\partial_x\alpha-\partial_x\lambda)+\beta^x\partial_x\Psi\right]\right\}\\
 &-\Big\{X,H_x\left[SN^x\partial_x\alpha\right]\Big\}+\Big\{X,H[\beta^x\partial_xN]\Big\}\\
 &+\Big\{X,H[\mathcal L_{\mathfrak N}(\alpha N)]+H_x[\mathcal L_{\mathfrak N}\beta^x]+\mathcal G[\mathcal L_{\mathfrak N}(\alpha\Psi+\lambda)]\Big\}.
 \end{aligned}
\end{equation}
Putting Eqs.~\eqref{eq:result1} and \eqref{eq:result2} into Eq.~\eqref{eq:dPG} yields
\begin{equation}\label{eq:finalresult}
 \begin{aligned}
 &\left\{X,H_x\left[\delta\mathfrak N^x-SN^2\partial_x\alpha-(\mathcal L_{\beta}\mathfrak N)^x\right]\right\}\\
 &+\Big\{X,H[\beta^x\partial_xN+\mathcal L_{\mathfrak N}(\alpha N)-\delta N]\Big\}\\
 &+\left\{X,\mathcal G\left[\xi\right]\right\}=0,
 \end{aligned}
\end{equation}
where the coefficient $\xi$ is
\begin{equation}
 \xi=-N \sqrt{S}(\Psi\partial_x\alpha+\partial_x\lambda)+\beta^x\partial_x\Psi+\mathcal L_{\mathfrak N}(\alpha\Psi+\lambda)-\delta\Psi,
\end{equation}
Note that the first two terms in Eq.~\eqref{eq:finalresult} vanish due to Eq.~\eqref{eq:dNNxEM}. Thus, Eq.~\eqref{eq:finalresult} gives
\begin{equation}
 \delta\Psi=-N\sqrt{S}(\Psi\partial_x\alpha+\partial_x\lambda)+\beta^x\partial_x\Psi+\mathcal L_{\mathfrak N}(\alpha\Psi+\lambda).
\end{equation}

\section{Detained calculations on Eq.~\eqref{eq:effectiveconstraintalge}}\label{appendix:B}

Since the Gauss constraint $\mathcal G$ contains only $P_\Gamma$, Eqs.~\eqref{eq:GG}, \eqref{eq:GHx}, and \eqref{eq:GH} can easily be checked. For Eqs.~\eqref{eq:HxHx-eff} and \eqref{eq:HxH-eff}, the results are just a direct consequence that $H_x$ generates spatial diffeomorphism transformation. Specifically, using the expression of $H_x=H_x^{GR}+H_x^{EM}$ given in Eqs.~\eqref{eq:HxGR} and \eqref{eq:HxEM}, we have
\begin{equation}\label{eq:diff1}
 \begin{aligned}
 \{O,H_x[N^x]\}=&N^x\partial_x O=\mathcal L_{N^x\partial_x} O, \forall O=K_2,E^1,P_\Gamma,\\
 \{\tilde O,H_x[N^x]\}=&\partial_x(N^x\tilde O)=\mathcal L_{N^x\partial_x} \tilde O, \forall \tilde O=E^2,K_1,\Gamma, \partial_xP_\Gamma.
 \end{aligned}
\end{equation}
where we should note that $\tilde O$ are scalar density with weight 1. As a consequence of these equations, we have
\begin{equation}\label{eq:diff2}
 \{s_a,H_x[N^x]\}=N^x\partial_x s_a=\mathcal L_{N^x\partial_x} s_a, \forall a=3,4,5,
\end{equation}
where the second equality results from that $s_a$ are scalars. According to Eqs.~\eqref{eq:diff1} and \eqref{eq:diff2}, for any phase space function $F$ depending on $v^I\in \{P_\Gamma, \Gamma, E^2, K_1, s_1\equiv E^1,s_2\equiv K_2,s_3,s_4,s_5\}$, we get
\begin{equation}
 \{F,H_x[N^x]\}=\frac{\partial F}{\partial v^I}\{v^I,H_x[N^x]\}=\mathcal L_{N^x\partial_x}F.
\end{equation}
Clearly, $H_x$ and $H_{\rm eff}$ are such a functions. Thus,we get
\begin{equation}
 \begin{aligned}
 \{H_x[N_1^x],H_x[N_2^x]\}=&\int \dd xN_1^x\mathcal L_{N_2^x\partial_x}H_x\\
 =&\int\dd xN_1^x\left(N_2^x\partial_xH_x+2H_x\partial_xN_2^x\right),\\
 \{H_x[N^x],H_{\rm eff}[N]\}=&-\int \dd xN\mathcal L_{N^x\partial_x}H_{\rm eff}\\
 =&-\int \dd xN\partial_x\left(N^xH_{\rm eff}\right),
 \end{aligned}
\end{equation}
which is the same as Eqs.~\eqref{eq:HxHx-eff} and \eqref{eq:HxH-eff} up to a boundary term. The only remaining equation that requires further effort is Eq.~\eqref{eq:HH-eff}.

For Eq.~\eqref{eq:HH-eff}, due to
\begin{equation*}
 \{O[N_1],O[N_2]\}=H_x[\mu S(N_1\partial_xN_2-N_2\partial_xN_1)],\, \forall O=H_{\rm eff}^{GR},\, H_{\rm eff}^{EM},
\end{equation*}
we only need to show the cross term in $\{H_{\rm eff}[N_1],H_{\rm eff}[N_2]\}$ vanishes. Namely, we need to show
\begin{equation}
 \{H_{\rm eff}^{GR}[N_1],H_{\rm eff}^{EM}[N_2]\}+\{H_{\rm eff}^{EM}[N_1],H_{\rm eff}^{GR}[N_2]\}=0.
\end{equation}
Observing that $H_{\rm eff}^{GR}$ does not depend on $\Gamma$ and $P_\Gamma$, we have, according to Eq.~\eqref{eq:hemeff} which indicates that $H_{\rm eff}^{EM}$ is a function of $\mu$, $E^I$, $\Gamma$, $P_\Gamma$, and $\partial_xP_\Gamma$,
\begin{equation}\label{eq:B6}
 \begin{aligned}
 &\{H_{\rm eff}^{GR}[N_1],H_{\rm eff}^{EM}[N_2]\}+\{H_{\rm eff}^{EM}[N_1],H_{\rm eff}^{GR}[N_2]\}\\
 =&\sum_{I=1,2}\int\dd xN_2(x)\frac{\partial H_{\rm eff}^{EM}}{\partial E^I}\left\{H_{\rm eff}^{GR}[N_1],E^I\right\}\\
 &+\int\dd xN_2(x)\frac{\partial H_{\rm eff}^{EM}}{\partial \mu}\left\{H_{\rm eff}^{GR}[N_1],\mu\right\}\\
 &-N^1\leftrightarrow N^2.
 \end{aligned}
\end{equation}
The notation $N_1 \leftrightarrow N_2$ indicates that the final line is obtained by exchanging $N_1$ and $N_2$ in the preceding terms. Since $H_{\rm eff}^{GR}$ does not depend on the derivatives of $K_I$ for $I=1,2$, we have, for the first term in the right hand side of Eq.~\eqref{eq:B6},
\begin{equation}
 N_2(x)\frac{\partial H_{\rm eff}^{EM}}{\partial E^I}\left\{H_{\rm eff}^{GR}[N_1],E^I\right\}\propto N_1(x)N_2(x),
\end{equation}
which will be exactly canceled by its counterpart in the antisymmetric combination appearing in the expression with $N_1 \leftrightarrow N_2$. This result leads us to simplify Eq.~\eqref{eq:B6} to be
\begin{equation}\label{eq:HGREM1}
 \begin{aligned}
 &\{H_{\rm eff}^{GR}[N_1],H_{\rm eff}^{EM}[N_2]\}+\{H_{\rm eff}^{EM}[N_1],H_{\rm eff}^{GR}[N_2]\}\\
 =&\int\dd xN_2(x)\frac{\partial H_{\rm eff}^{EM}}{\partial \mu}\left\{H_{\rm eff}^{GR}[N_1],\mu\right\}-N^1\leftrightarrow N^2\\
 =&\left(\{H_{\rm eff}^{GR}[N_1],A[\mu]\}-N_1\leftrightarrow A\right)\Big|_{A=N_2(x)\frac{\partial H_{\rm eff}^{EM}}{\partial \mu}},
 \end{aligned}
\end{equation}
where $A$ denotes a phase-space-independent scalar density. We thus need to show $\{H_{\rm eff}^{GR}[N_1],A[\mu]\}-N_1\leftrightarrow A$ vanishes. Noting that $\mu$ depends on $s_1$, $s_2$, and $s_4$, we have
\begin{equation}\label{eq:Hmu}
 \begin{aligned}
 &\{H_{\rm eff}^{GR}[N_1],A[\mu]\}-N_1\leftrightarrow A\\
 =&-\int\dd x\dd yA(x)N_1(y)(\partial_{s_2}\mu)(x) \left(\partial_{s_5} H_{\rm eff}^{GR}\right)(y)\{s_2(x),s_5(y)\}\\
 &+\int\dd x\dd yN_1(x)A(y)\left(\partial_{s_3} H_{\rm eff}^{GR}\right)(x)(\partial_{s_4}\mu)(y)\{s_3(x),s_4(y)\}\\\\
 &-N_1\leftrightarrow A,
 \end{aligned}
\end{equation}
where we have omitted terms involving Poisson brackets of the form $\{s_a(y), s_b(x)\} \propto \delta(x, y)$, as these lead to contributions proportional to $A(x) N_1(x)$ and thus are canceled under the antisymmetrization $N_1 \leftrightarrow A$. A direct calculation yields
\begin{equation}
 \begin{aligned}
 \{s_2(x),s_5(y)\}&=\left[3G\frac{[\partial_yE^1(y)][\partial_yE^2(y)]}{[E^2(y)]^4}-2G\frac{\partial_y^2E^1(y)}{[E^2(y)]^3}\right]\delta(x,y)\\
 &\quad-G\frac{\partial_yE^1(y)}{[E^2(y)]^3}\partial_y\delta(x,y),
 \end{aligned}
\end{equation}
and
\begin{align}
 \{s_3(x),s_4(y)\}&=\frac{2G}{E^2(x)E^2(y)}\partial_y\delta(x,y).
\end{align}
Substituting the results into \eqref{eq:Hmu}, we get
\begin{equation}\label{eq:Hmu1}
 \begin{aligned}
 &\{H_{\rm eff}^{GR}[N_1],A[\mu]\}-N_1\leftrightarrow A\\
 =&-G\int\dd x \left(A\partial_xN_1-N_1\partial_xA\right) \frac{\partial_xE^1(\partial_{s_2}\mu)\partial_{s_5} H_{\rm eff}^{GR}}{[E^2]^3}\\
 &-G\int\dd x\dd y\left(N_1 \partial_xA -A \partial_xN_1 \right)\frac{2\left(\partial_{s_3} H_{\rm eff}^{GR}\right) (\partial_{s_4}\mu)}{[E^2]^2},
 \end{aligned}
\end{equation}
According to Eqs.~\eqref{eq:covariance-Heff} and \eqref{eq:covarianceequationgeneral2}, we have
\begin{equation}
 \frac{\partial_xE^1(\partial_{s_2}\mu)\partial_{s_5} H_{\rm eff}^{GR}}{[E^2]^3}=\frac{-2(\partial_{s_2}\mu)( \partial_{s_4}M_{\rm eff})}{E^2}=\frac{2\left(\partial_{s_3} H_{\rm eff}^{GR}\right) (\partial_{s_4}\mu)}{[E^2]^2}.
\end{equation}
Inserting this result into Eq.~\eqref{eq:Hmu1} leads to that $\{H_{\rm eff}^{GR}[N_1],A[\mu]\}-N_1\leftrightarrow A$ vanishes, as what we wanted.



\end{document}